\documentclass[12pt,a4paper]{article}
\pdfoutput=1

\usepackage{graphicx}
\usepackage{caption}
\usepackage{subcaption}
\usepackage{jheppub}
\usepackage{enumitem}
\usepackage{framed,float}
\usepackage{booktabs}
\usepackage{subcaption}
\usepackage{tikz}
\usetikzlibrary{calc,arrows,decorations.markings}
\usepackage{mathrsfs}
\usepackage[utf8]{inputenc}

\usepackage{color}

\def\CO{{\cal O}}
\def\IC {{\mathbb C}}
\def\IE {{\mathbb E}}
\def\IF {{\mathbb F}}
\def\IN{{\mathbb N}}
\def\IP{{\mathbb P}}
\def\IQ{{\mathbb Q}}
\def\IR {{\mathbb R}}

\def\IZ {{\mathbb Z}}

\newtheorem{thm}{Theorem}[section]
\newtheorem{prop}[thm]{Proposition}
\newtheorem{cor}[thm]{Corollary}
\newtheorem{defn}[thm]{Definition}
\newtheorem{claim}[thm]{Claim}

\title{Swampland Constraints on 5d $\mathcal{N}=1$ Supergravity}

\author[a]{Sheldon Katz,\,}
\author[b,c]{Hee-Cheol Kim,\,}
\author[d]{Houri-Christina Tarazi,\,}
\author[d]{Cumrun Vafa\,}

\affiliation[a]{Department of Mathematics, MC-382, University of Illinois at Urbana-Champaign, Urbana, IL 61801, USA}
\affiliation[b]{Department of Physics, POSTECH, Pohang 790-784, Korea}
\affiliation[c]{Asia Pacific Center for Theoretical Physics, Postech, Pohang 37673, Korea}
\affiliation[d]{Department of Physics, Harvard University, Cambridge, MA 02138, USA}

\abstract
{
We propose Swampland constraints on consistent 5-dimensional ${\cal N}=1$ supergravity theories.  We focus on a special class of BPS magnetic monopole strings which arise in gravitational theories.   The central charges and the levels of current algebras of 2d CFTs on these strings can be calculated by anomaly inflow mechanism and used to provide constraints on the low-energy particle spectrum and the effective action of the 5d supergravity based on unitarity of the worldsheet CFT.  In M-theory, where these theories are realized by compactification on Calabi-Yau 3-folds, the special monopole strings arise from wrapped M5-branes on special (``semi-ample'') 4-cycles in the threefold.  We identify various necessary geometric conditions for such cycles to lead to requisite BPS strings and translate these into constraints on the low-energy theories of gravity.  These and other geometric conditions, some of which can be related to unitarity constraints on the monopole worldsheet, are additional candidates for Swampland constraints on 5-dimensional ${\cal N}=1$ supergravity theories.
}

\begin{document}
\begin{flushright}
\tt 
\end{flushright}

\maketitle

\section{Introduction}\label{sec:intro}
Many decades of string theory research has reinforced the picture that quantum field theories that arise in the low energy limit of quantum gravitational theories are rather special \cite{Vafa:2005ui}.  The conditions thus imposed from string theory on the low energy quantum gravity theories, the Swampland constraints, are expected to cut the space of quantum field theories (up to deformations) to a finite set.    However, a skeptic may wonder whether string theory's inability to lead to specific low energy quantum systems is a deficiency of string theory, or a constraint on quantum gravitational theories.
In other words, could the string lamppost be misleading us in identifying the correct Swampland constraints?  Or does the string lamppost principle (that all consistent quantum gravitational theories are accounted for by the string vacua) hold?

To investigate this question we need to find reasons, independently of string theory, which explain the proposed Swampland constraints based on other basic principles, such as unitarity and other consistency requirements of quantum gravity theories.  To study this question it would be natural to start with the simplest class of quantum gravity theories with the higher amount of supersymmetries in Minkowski background.  Theories with 32 supercharges are unique in each dimension (except there are two in 10 dimensions depending on whether we have IIA $(1,1)$ or IIB $(2,0)$ supersymmetries), and all realized in string theory.   So the next case to consider are theories with 16 supercharges.  There are only 2 such theories in 10 dimensions and that can be argued based on anomalies combined with additional arguments \cite{Adams:2010zy,Kim:2019vuc} which rule out additional possibilities with abelian factors.  Again both of these theories ($E_8\times E_8$ and $SO(32)$) are obtained from string theory.  Theories with 16 supercharges in lower dimensions obtained from string theory are rather restricted and in particular they enjoy an upper bound on their rank $r\leq 26-d$. This upper bound on the rank has also been obtained recently without appealing to string theory \cite{Kim:2019ths}.  Given this success in checking the string lamppost principle with theories with such a high supersymmetry, it is natural to move to the next case of theories with 8 supercharges\footnote{We can also study theories with 24 and 12 supercharges as well, which arise in $d\leq 4$.}.

The highest dimension for supergravity theories with 8 supercharges is $d=6$, corresponding to ${\cal N}=(1,0)$ supersymmetric theories.   Being chiral, anomaly cancellations lead to severe restrictions on the matter content of these theories, which has been systematically studied \cite{Kumar:2009ac}.  A finite subset of the anomaly free matter contents can be realized by compactification of F-theory on elliptic CY 3-folds and it is natural to believe that the rest are not consistent.  In \cite{Kim:2019vuc} it was shown that the assumption of existence of BPS strings in these theories (which is a natural extension of Swampland's completeness conjecture), and the requirement of having a unitary theory on their worldsheet, leads to severe restrictions and it was shown that this rules out some infinite class of anomaly free matter spectrum that is not possible to construct from string theory.  The idea of using unitarity of BPS worldsheets in the context of swampland has been further investigated in \cite{Lee:2019skh,Kim:2019ths}.

The aim of this paper is to continue investigating Swampland constraints on theories with 8 supercharges along the same lines, but now in 5 dimensions.  These are obtained by compactification of M-theory on CY 3-folds.  A subset of these involve compactification on elliptic CY 3-folds, which would correspond to further compactification of 6d F-theory constructions of ${\cal N}=(1,0)$ supersymmetric theories on a circle.
In addition to the matter content, the low energy theory is characterized by gauge ($AF^2$) as well as mixed gauge/gravitational Chern-Simons terms ($AR^2$).  Our aim in this paper is therefore to find restrictions on both the matter content and the associated Chern-Simons terms.   5d supergravities enjoy BPS strings, which in the gauge theory setup would correspond to monopole strings.  We focus on a special class of these strings which exist only in gravitational theories and which give rise to black strings with macroscopic entropy when we increase their charge.    As in \cite{Kim:2019vuc} we use unitarity constraints on the worldsheet of monopole strings to find Swampland constraints.
From the M-theory perspective, these strings correspond to M5 branes wrapping special 4-cycles (``semi-ample''), which roughly speaking translate to the condition that the cycles are represented by holomorphic cycles which are not rigid.  These are a slight extension of MSW strings studied in \cite{Maldacena:1997de}.   In this geometric setup, the gauge and mixed Chern-Simons terms translate to triple intersection of 4-cycles and to the intersection of 4-cycle with dual to second Chern class, respectively.  We use this dictionary to see which constraints predicted from the unitarity constraints on the monopole worldsheet can be seen geometrically.  We find that some, but not all of them can be explained geometrically.  Similarly, from geometric restriction, we find additional constraints for the effective 5d supergravity theories that we do not know their origin in terms of unitarity constraints on the monopole strings.  We propose the union of these constraints as candidates for Swampland constraints for ${\cal N}=1$ supergravity theories in 5 dimensions.

The rest of the paper is organized as follows. In Section \ref{sec:2} we review the 5d $\mathcal{N}=1$ supergravity theories. The Coulomb branch of the moduli space and the effective action, and BPS monopole strings in the supergravity theories will be introduced. We also discuss the consistency conditions of monopole strings in terms of Chern-Simons terms in the 5d effective action. In Section \ref{sec:Mtheory}, we identify monopole strings in the low-energy 5d theories in M-theory compactified on a compact Calabi-Yau threefold with M5-brane states wrapped on semi-ample divisors embedded in the threefold. We give the details of geometric conditions on semi-ample divisors and discuss the physical interpretations of the geometric conditions.
In Section \ref{sec:constraints}, we show that the conditions on the strings and the properties of the associated 4-cycles when geometrically engineered can be used to constrain the effective Chern-Simons terms and the content of massless states in the 5d gravity theories.
Lastly, in Section \ref{sec:conclusion} we present our conclusions and discuss some open questions. Appendix \ref{appendix} contains various mathematical facts about the 4-cycles in Calabi-Yau threefolds. We provide in Appendix \ref{app:examples} explicit constructions of some Calabi-Yau geometries that lead to 5d supergravity theories.

\section{ $\mathcal{N}=1$ supergravity in five dimensions}\label{sec:2}
In this section, we start by reviewing the salient features of the $\mathcal{N}=1$ supergravity theory in five dimensions. We will first present a gauge theoretic perspective of the 5d supergravity theory that is described by the effective theory of gravity coupled to vector multiplets for the gauge group $\mathcal{G}$ and hypermultiplets carrying the gauge charges. We will then review the monopole strings of the 5d theory and then define a class of those called  \textit{supergravity strings} which exist only in a supergravity theory.

\subsection{General aspects of  low energy field theory}\label{sec:fieldtheory}
Consider a 5d $\mathcal{N}=1$ gravitational theory with gauge group $\mathcal{G}$. 
We are primarily interested in its effective field theory at low energy on the Coulomb branch of the moduli space. The massless supermultiplets in the spectrum are the gravity multiplet, a number of vector multiplets, and charged and neutral hypermultiplets.
The vector multiplets of $\mathcal{G}$ contain the vector fields $A_\mu$ and the real scalar fields $\phi$. The scalars $\phi$ can take nonzero expectation values in the Cartan subalgebra of the gauge group $\mathcal{G}$. The scalar expectation values, which we denote by $\phi^a$, $a=1,\cdots,\mathbf{r}$, are moduli parametrizing the Coulomb branch.
At a generic point on the Coulomb branch, the gauge group $\mathcal{G}$ is broken to its abelian subgroup $U(1)^{\mathbf{r}}$ with $\mathbf{r}={\rm rank}(\mathcal{G})$ and the theory reduces to a supergravity theory coupled to $\mathbf{r}$ Abelian vector multiplets as well as neutral hypermultiplets.

The bosonic action on the Coulomb branch for the gravity multiplet and vector multiplets is given by \cite{Ceresole:2000jd,Bonetti:2011mw}
\begin{equation}
	S=\int\left( *R - G_{IJ}d \phi^I\wedge *d \phi^J-G_{IJ}F^I\wedge *F^J - \frac{1}{6}C_{IJK}A^I \wedge F^J \wedge F^K \right) \ ,
\end{equation}
where $R$ is the Ricci curvature and $F^I=dA^I$ is the field strength of the gauge group. Here we collectively denote the graviphoton field $A^0_\mu$ in the gravity multiplet and the $\mathbf{r}$ gauge fields in the vector multiplets by $A^I_\mu$, $I=0,1,\cdots \mathbf{r}$. $G_{IJ}$ is the metric for the geometry of the scalar moduli space. $C_{IJK}$ is the level for the cubic Chern-Simons term and it is quantized due to gauge invariance of the Abelian symmetries as $C_{IJK}\in \mathbb{Z}$ \cite{Witten:1996qb}.

The metric on the scalar moduli space in the effective action is determined by the prepotential defined as
\begin{equation}
	\mathcal{F} = \frac{1}{6}C_{IJK}\phi^I\phi^J\phi^K \ ,
\end{equation} 
which is a homogeneous cubic polynomial in the scalar expectation values $\phi^I$ obeying the hypersurface constraint,
\begin{equation}
	\mathcal{F} = \frac{1}{6}C_{IJK}\phi^I\phi^J\phi^K  = 1 \ .
\end{equation}
The geometry parametrized by $\phi^I$ under this constraint is called the {\it very special geometry}.
The metric on this hypersurface can be obtained as
\begin{equation}\label{eq:metric}
	G_{IJ} = -\left.\frac{1}{2}\frac{\partial^2\log \mathcal{F}}{\partial \phi^I \partial \phi^J}\right|_{\mathcal{F}=1} \ .
\end{equation}

One can also consider hypermultiplets in the low energy theory. Charged hypermultiplets are all massive at a generic point on the Coulomb branch. They are already integrated out in the above effective action. On the other hand, the neutral hypermultiplets remain massless in the low energy theory. They will play some role in our discussion later.

On the Coulomb branch of the moduli space, the low-energy spectrum includes 1/2 BPS extended objects magnetically charged under the Abelian gauge groups. 
We call them magnetic monopole strings, or monopole strings for short. A monopole string carries the magnetic charge of gauge fields $A^I$ as,
\begin{equation}
	q^I = \frac{1}{2\pi}\int_{S^2}F^I \ .
\end{equation}
Here the integration is taken over the two-sphere $S^2$ surrounding the string. 
The tension of this string can be exactly computed from the prepotential. For a string carrying unit magnetic charge of $A^I_\mu$, its tension is given by
\begin{equation}\label{eq:Tension}
	T_I = \left.\frac{\partial \mathcal{F}}{\partial \phi^I} \right|_{\mathcal{F}=1} \ .
\end{equation}
The monopole strings and their geometric counterparts are central ingredients in this paper. We will discuss their properties in detail in the next subsections.

The low-energy theory may involve higher derivative corrections to the effective action. Some special higher derivative terms are determined by a combination of topological data and supersymmetry.
One is the mixed gauge/gravitational Chern-Simons term of the form,
\begin{equation}
	S_{ARR} = \frac{1}{96}\int C_I\,A^I\wedge {\rm tr}( R\wedge R) \ ,
\end{equation}
where $R=d\omega +\omega \wedge \omega$ is the curvature 2-form for the spin connection $\omega$. This term is linear in the gauge field $A^I$, so we call this term as the linear Chern-Simons term. 
The supersymmetric completion of this four-derivative correction was obtained in \cite{Hanaki:2006pj} using  conformal supergravity techniques.

The level $C_I$ for the linear Chern-Simons term is also quantized as follows \cite{Chang:2019uag}. Let us put the theory on a five manifold $\mathcal{M}_5 = S^1\times\mathcal{M}_4$. Then consider a large gauge transformation of the gauge field $A^I$,
\begin{equation}
	A^I \ \rightarrow \ A^I+\frac{n}{R}dx^5 \ ,
\end{equation}
where $n\in \mathbb{Z}$ and $R$ is the radius of the $S^1$ with coordinate $x^5$. This gauge transformation varies the linear Chern-Simons term as
\begin{equation}
	\delta S_{ARR} = -\frac{n\pi}{24} C_I \int_{\mathcal{M}_4} p_1(T_4) \ ,
\end{equation}
where $p_1(T_4)$ is the first Pontryagin class for the tangent bundle $T_4$. Note that the integration of $p_1$ over a spin manifold $\mathcal{M}_4$ gives an integer number:
\begin{equation}
	\frac{1}{48}\int_{\mathcal{M}_4}p_1(T_4) \in \mathbb{Z} \ .
\end{equation}
Now demanding that the partition function is invariant under this large gauge transformation quantizes the level as an even integer, therefore $C_I \in \mathbb{Z}$.

The gauge symmetry can enhance to a bigger symmetry at special loci in the moduli space where some charged vector fields become massless. On the special locus, the Abelian gauge groups can enhance to non-Abelian groups provided that the massless charged vector fields form the adjoint representation of the non-Abelian symmetries.
The full gauge group is then given by $\mathcal{G}=G\times U(1)^{\mathbf{r}+1-r}$ where $G$ is the product of the enhanced non-Abelian groups with $r={\rm rank}(G)$. There can also be massless hypermultiplets charged under the enhanced gauge symmetry $G$. The low-energy theory on the special vacua is then described by the gauge theory of the enhanced gauge group $\mathcal{G}$ coupled to the massless charged hypermultiplets.

Two-derivative terms in the gauge theory action for each gauge multiplet $\Phi_i$ of a simple non-Abelian group $G_i\subset G$ are determined from the prepotential
 \begin{equation}
 	\mathcal{F}_{G_i} =- \frac{h_{i} }{2}{\rm Tr}(\Phi_i^2) + \frac{\kappa_{i}}{6}{\rm Tr}(\Phi_i^3) \ ,
 \end{equation}
where $h_i$ is the gauge coupling and $\kappa_i$ is the classical Chern-Simons level for $G_i$. The classical Chern-Simons level is an integer and non-zero only for $G_i=SU(N)$ with $N\ge3$. Here, the gauge coupling $h_i$ is given by a linear sum of the scalar values $\phi^\alpha$ in the Abelian part, such as
\begin{equation}
	h_i=\sum_{\alpha=1}^{\mathbf{r}+1-r}h_{i,\alpha}\phi^\alpha \ .
\end{equation}
These scalar moduli $\phi^\alpha$ parametrize the special sub-manifold of the moduli space where the gauge symmetry enhancement occurs.
There is no four-derivative correction to the non-Abelian action because if it exists, it is linear in $\Phi_i$, but $\Phi_i$ is traceless.

One can move away from the special vacua by turning on generic scalar expectation values, say $\phi_i^a$, $a=1,\cdots,{\rm rank}(G_i)$, for the Cartan generators of the non-Abelian symmetry $G_i$. This will bring us back to the Abelian effective theory at low energy. The prepotential of the Abelian theory in the neighborhood of the special loci is determined by a one-loop calculation with charged fermions that become massive with non-zero values of $\phi_i^a$. For a non-Abelian gauge group $G_i$ and matter hypermultiplets in generic representations, the prepotential after the one-loop calculation is \cite{Witten:1996qb,Seiberg:1996bd,Intriligator:1997pq}
\begin{equation}
	\mathcal{F}_{G_i} =- \frac{h_i}{2} \,K_{i,ab}\, \phi^a_i \phi^b_i +\frac{\kappa_i}{6} d_{i,abc}\phi^a_i\phi^b_i\phi^c_i + \frac{1}{12}\left(\sum_{\mathbf{R}}|\mathbf{R}\cdot \phi_i|^3 -\sum_f\sum_{\mathbf{w}_f}|\mathbf{w}_f\cdot \phi_i|^3\right) \ ,
\end{equation}
where $K_{i,ab}$ is the Killing form of $G_i$ and $d_{i,abc}=\frac{1}{2}{\rm Tr}(T^a_i\{T^b_i,T^c_i\})$ with the generator $T^a_i$ in the fundamental representation of $G_i$. $\mathbf{R}$ and $\mathbf{w}_f$ are the roots and the weights for the $f$-th hypermultiplet of $G_i$, respectively.

In addition, a mixed gauge/gravitational Chern-Simons term with the level $C_{i,a}$ is induced by integrating out the charged fermions. The result from the one-loop computation is \cite{Bonetti:2013ela}
\begin{equation}
	C_{i,a} = - \frac{\partial}{\partial \phi^a_i}\left( \sum_{\mathbf{R}}|\mathbf{R}\cdot\phi_i| - \sum_f \sum_{\mathbf{w}_f}|\mathbf{w}_f\cdot \phi_i| \right)\ .
\end{equation}

Note that not all supergravity theories have such special sub-manifolds of the moduli space supporting enhanced gauge symmetry. Also, it is possible that a single theory has different special vacua with different non-Abelian gauge theory descriptions, which may lead to interesting dualities.
However, at a generic point on the Coulomb branch, the effective theory after integrating out all the massive charged fields always reduces to an Abelian gauge theory description.

A large class of 5d $\mathcal{N}=1$ supergravities can be constructed from M-theory compactification on compact Calabi-Yau threefolds. Such theories will be discussed in more details in the next section.

\subsection{Monopole strings in 5d supergravities}\label{sec:Monopole}
As we reviewed in the previous subsection, 5d supergravity theories contain magnetic monopole strings. Here we would like to study their basic properties. Monopole strings are two-dimensional magnetic sources for the low-energy Abelian gauge fields on the Coulomb branch.  In particular, we shall consider 1/2 BPS monopole string configurations preserving 4 chiral supercharges in the 2d worldsheet.

The monopole string with magnetic charge $q^I$ can be introduced by a delta function source in the Bianchi identity of the gauge field strength $F^I=dA^I$ as,
\begin{equation}\label{eq:string-source}
	dF^I =q^I \prod_{\mu=2}^4\delta(x^\mu)dx^\mu \ .
\end{equation}
We assume here that the monopole source is located at the origin $x^{2,3,4}=0$ on the transverse $\mathbb{R}^3$. 
Completeness of charged string spectrum in a gravitational theory ensures existence of such monopole string states as long as Dirac quantization condition is obeyed \cite{Polchinski:2003bq,Banks:2010zn} and the string tension is positive.
In the following discussions, we shall focus on single BPS monopole string states for a given primitive magnetic charge $q^I$. The question of existence of such string states will be discussed later.

The string source supports a microscopic 2d theory that flows in the IR to a 2d $\mathcal{N}=(0,4)$ SCFT.
The worldsheet SCFT involves chiral degrees of freedom coming from zero modes of the charged fields in the bulk gauge theory on the string background. The 2d chiral fields charged under the bulk symmetry develop non-trivial anomalies for the symmetries. The anomaly arising from the worldsheet degrees of freedom must be cancelled by an other source since otherwise, the monopole string configuration in the 5d supergravity will be inconsistent by the quantum anomaly along the string worldsheet.

The anomaly cancellation can be achieved by the anomaly inflow mechanism from the bulk gravity theory toward the string source. 
The anomaly inflow in the presence of BPS monopole strings in 5d supersymmetric theories was studied in \cite{Ferrara:1996hh,Mizoguchi:1998wv,Boyarsky:2002ck} (See also \cite{Callan:1984sa,Berman:2004ew,Henningson:2004dh,Kim:2016foj,Shimizu:2016lbw,Kim:2019vuc,Kim:2019ths} for anomaly inflow of BPS strings in other dimensions). We shall generalize these earlier studies and compute the anomaly inflow in the presence of the string sources in 5d supergravities. Using the result we will then compute gravitational and 't Hooft anomalies as well as central charges of the 2d SCFTs on monopole strings.

Let us first compute the anomaly inflow induced from the bulk Chern-Simons terms. The Chern-Simons terms in the bulk effective action is no longer invariant under the symmetry transformations when monopole strings are introduced. 

When the string source with magnetic charge $q^I$ in (\ref{eq:string-source}) is inserted, the cubic Chern-Simons term transforms under the local gauge transformation $\delta A^I = d\Lambda^I$ as \cite{Ferrara:1996hh},
\begin{eqnarray}
	\delta_{\Lambda} S_{\rm cs} &=& \int_{M_5}\left(-\frac{1}{2}C_{IJK} d\Lambda^I\wedge F^J \wedge F^K \right) = C_{IJK} \int_{M_5} \Lambda^I F^J \wedge dF^K \nonumber \\
	&=& C_{IJK}q^K\int_{M_2} \Lambda^I F^J \ .
\end{eqnarray}
We here used the modified Bianchi identity in (\ref{eq:string-source}) for the second line. Thus the gauge variation does not vanish for general charge and Chern-Simons level.
This non-vanishing gauge anomaly is the gauge anomaly inflow induced along the 2d string worldsheet.

The gravitational anomaly inflow computation for the local Lorentz transformation is more involved. In particular both the cubic and the linear Chern-Simons terms contribute to the gravitational anomaly inflow. To compute these contributions, we first solve the Bianchi identity in (\ref{eq:string-source}) of a string source by using the magnetic flux of the smoothed form \cite{Freed:1998tg,Boyarsky:2002ck}
\begin{equation}
	F^I = -\frac{1}{2}q^I  d\rho \wedge e_1^{(0)} \ ,
\end{equation}
where $\rho(r)$ is a smooth function of the radial direction $r$, with $\rho(0)=-1$ and $\rho(r)=0$ for sufficiently large $r$, and $e_1^{(0)}$ is the 1-form in the descent relations $de^{(0)}_1 = e_2,\,\delta e_1^{(0)}=de_0^{(1)}$ for the global angular form $e_2$ of the 2-sphere surrounding the monopole string. This smooths out the string source as
\begin{equation}
	dF^I = q^I d(\rho e_2/2) \ .
\end{equation}
In this case, the gauge field for the magnetic flux transforms under diffeomorphisms as $\delta A^I = -\frac{1}{2}q^I d(\rho e^{(1)}_0)$. The following integrals for the 2-form $e_2$ on the 2-sphere bundle over the string worldvolume will prove to be useful for later discussion:
\begin{equation}
	\int_{S^2}e_2 = 2 \ , \quad \int_{S_\epsilon(M_2)}e_0^{(1)}e_2\wedge e_2 = 2\int_{M_2}p_1^{(1)}(N) \ ,
\end{equation}
where $p_1^{(1)}(N)$ is the 2-form in the descent relation, $dp_1^{(0)}(N)=p_1(N)$ and $\delta p_1^{(0)}(N)= dp_1^{(1)}(N)$, of the first Pontryagin class $p_1(N)$ of the $SU(2)_R$ normal bundle for the transverse $\mathbb{R}^3$ directions.

We now consider the local Lorentz transformation of the effective action on the background magnetic flux. One can compute the variation of both the cubic and the linear Chern-Simons terms under the local Lorentz transformation as \cite{Freed:1998tg,Boyarsky:2002ck}
\begin{eqnarray}
	\delta_g S_{\rm cs} &=& \frac{1}{48}C_{IJK}q^Iq^Jq^K \int_{M_5} d(\rho e_0^{(1)})e_2^2 +\frac{1}{96}C_Iq^I\int_{M_5} e_2 \wedge \delta p_1^{(0)}(T_5) \nonumber \\
	&=& -\frac{1}{24}C_{IJK}q^Iq^Jq^K\int_{M_2}p_1^{(1)}(N) -\frac{1}{48}C_Iq^I\int_{M_2}p_1^{(1)}(T_5) \ ,
\end{eqnarray}
where $p_1(T_5)$ is the first Pontryagin class of the tangent bundle $T_5$ of the 5d spacetime. 
This non-vanishing variation of the bulk action is the gravitational anomaly inflow toward the monopole string.

The anomaly inflow for the gauge and the Lorentz transformations must be cancelled by the anomalies developed by the worldsheet degrees of freedom living on the monopole strings. This fact allows us to compute the quantum anomaly of the 2d CFT on the string worldsheet from the anomaly inflow that we just computed. Collecting the above results, we conclude that the 2d SCFT on the monopole string with magnetic charge $q^I$ must have gauge and gravitational anomalies that are encoded in the 4-form anomaly polynomial of the form,
\begin{eqnarray}\label{eq:anomaly}
	I_4 &=& -I_4^{\rm inflow}  \\
	&=&-\frac{1}{2}C_{IJK}q^IF^J F^K+\frac{1}{24}C_{IJK}q^Iq^Jq^K p_1(N)+\frac{1}{48}C_Iq^Ip_1(T_5) \nonumber \\
	&=& -\frac{1}{2}C_{IJK}q^IF^J F^K -\frac{1}{6}\left(C_{IJK}q^Iq^Jq^K+\frac{1}{2}C_Iq^I\right)c_2(R) +\frac{1}{48}C_Iq^Ip_1(T_2) \ , \nonumber
\end{eqnarray}
where $I_4^{\rm inflow}$ is the anomaly inflow whose variation is related to the variation of the bulk action  $I_2^{(1)}=\delta S_{5d}$ via the descent relations $I_4^{\rm inflow}=d I_3,\, \delta I_3 = dI_2^{(1)}$. For the last line we used the relations of characteristic classes $p_1(T_5) = p_1(T_2) - 4c_2(R)$ and $p_1(N) = -4c_2(R)$, where $p_1(T_2)$ is the first Pontrygin class of the tangent bundle $T_2$ of the 2d worldsheet and $c_2(R)$ is the second Chern class of the $SU(2)_R$ Lorentz group transverse to the 2d worldsheet.

The anomaly polynomial $I_4$ of a 2d CFT encodes the left- and the right-moving central charges and the levels of the Kac-Moody current algebra coupled to the bulk gauge symmetry $\mathcal{G}$. The relative central charge $c_R-c_L$ can be read off from the coefficient of the gravitationaly anomaly term $-\frac{1}{24}p_1(T_2)$ in $I_4$. The right-moving central charge is $c_R = 6k_R$ where $k_R$ is 't Hooft anomaly coefficient of the $SU(2)$ R-symmetry in the IR $(0,4)$ superconformal algebra. In order to compute the individual left- and right-moving central charges of the IR CFT, we thus need to know the exact value of $k_R$, which demands us to identify the correct $SU(2)$ R-symmetry in the IR CFT.

\paragraph{Symmetry enhancement}
Naively, one expects that the $SU(2)_R$ symmetry would reduce to the IR R-symmetry of the 2d CFTs on monopole strings because this is the only $SU(2)$ symmetry under which the supercharges are charged. However this is not manifest in the cases when some accidental symmetry emerges in IR. 

For example, as we will describe in the next section there are monopole strings living on local 5d SCFTs which amount to M5-branes wrapping 4-cycles in local CY$_3$'s.
Such strings become tensionless strings in the CFT limit of the local theory when gravity decouples. The corresponding 4-cycles in a local CY$_3$ can collapse to zero size in the CFT limit.
 For those strings, the IR worldsheet CFT acquires an accidental $SU(2)_I$ symmetry inherited from the $SU(2)_I$ R-symmetry of the local 5d SCFT. This emergent $SU(2)_I$ symmetry, instead of $SU(2)_R$, in the IR CFT becomes the R-symmetry of the IR superconformal algebra. Therefore in this case the central charges should be calculated with respect to the $SU(2)_I$ symmetry. These strings can tell us physics of local 5d SCFTs. However since their low-energy physics is not affected by bulk gravitational interactions, we cannot use them to explore consistency of gravity theories. For this reason, we are not interested in these strings embedded in local 5d SCFTs with accidental $SU(2)_I$ symmetry.

Also, the strings arising from 6d self-dual strings by $S^1$ compactification have a different R-symmetry in their worldsheet CFTs at low-energy. The $SU(2)_R\times U(1)$ symmetry, where the $U(1)$ is for the KK momentum, in the worldsheet theory enhances to $SU(2)_r\times SU(2)_l$ in IR after decoupling the center-of-mass modes and the $SU(2)_r$, instead of $SU(2)_R$, becomes the R-symmetry of the IR $\mathcal{N}=(0,4)$ superconformal algebra in the interacting sector. Here $SU(2)_R$ is the diagonal subgroup of $SU(2)_l\times SU(2)_r$. The anomaly polynomial and the central charges of self-dual strings in 6d supergravities are computed in \cite{Kim:2019vuc} by using anomaly inflow mechanism (See also \cite{Kim:2016foj,Shimizu:2016lbw}). The anomaly polynomial of the 6d self-dual strings reduces to that of 5d monopole strings given in (\ref{eq:anomaly}) by identifying $q^I=Q^I$ and $C_I=-12a_I$, and also $c_2(l)=c_2(r)=c_2(R)$ from the relation $SU(2)_R\subset SU(2)_r\times SU(2)_l$ under $S^1$ reduction. From this, one can deduce that $C_{IJK}q^Iq^Jq^K=0$ and $C_Iq^I=-12Q\cdot a$ for the 6d self-dual strings. In M-theory compactified on CY3, the 6d self-dual strings correspond to M5-branes wrapped on elliptic surfaces equipped with elliptic fibration structure which will further be discussed in Section \ref{sec:geocon}. We note that when a 6d theory is compactified on a circle with automorphism twists, the worldsheet theory on a string that is affected by the twist do not have $SU(2)_l\times SU(2)_r$ symmetry enhancement since the Lorentz symmetry $SU(2)_l\times SU(2)_r$ is broken to $SU(2)_R \times U(1)$ by the twist. In this case, we expect that the $SU(2)_R$ will become the IR R-symmetry of the worldsheet CFTs.

It may also be possible that the IR worldsheet CFT shows supersymmetry enhancement. For example, 
the worldsheet CFT on self-dual strings in the 6d SCFT of $\mathcal{O}(-2)\rightarrow \mathbb{P}^1$ model is realized by a UV $\mathcal{N}=(0,4)$ gauge theory \cite{Haghighat:2013gba,Haghighat:2013tka}, but this theory is expected to flow in the infrared to a CFT with enhanced $\mathcal{N}=(4,4)$ supersymmetry. Another interesting example of 2d CFTs showing supersymmetry enhancement is the worldsheet theory on strings in the 9d supergravity theory constructed in M-theory on the Klein Bottle \cite{Dabholkar:1996pc}. This worldsheet theory naively has only $\mathcal{N}=(0,8)$ supersymmetry, but the IR SUSY turns out to get enhanced to $\mathcal{N}=(8,8)$ \cite{Aharony:2007du}.

Similarly, the $\mathcal{N}=(0,4)$ supersymmetry on monopole strings in 5d supergravity can also enhance to a larger SUSY in the infrared CFT.
Let us first discuss $\mathcal{N}=(4,4)$ SUSY enhancement.
In this case the enhanced superconformal algebra must be the small $\mathcal{N}=(4,4)$. The large $\mathcal{N}=4$ algebra in 2d CFTs involves two $SU(2)$ R-symmetries in each chiral sector. However, when coupled to 5d gravity we cannot have such two $SU(2)$ R-symmetries. So the enhanced $(4,4)$ symmetry can only be the small $\mathcal{N}=(4,4)$ symmetry. The small $\mathcal{N}=4$ conformal algebra involves a single (anti-)holomorphic $SU(2)$ R-symmetry which may be identified with the $SU(2)_R\subset SO(1,4)$ Lorentz symmetry in the 5d theory. However, the small $\mathcal{N}=(4,4)$ conformal algebra involves two copies of $\mathcal{N}=4$ conformal algebra referred to as the left-moving and right-moving sectors, and the $SU(2)$ R-symmetries in those two sectors are independent and distinct if the CFT is unitary and the vacuum is normalizable. We expect after removing the center-of-mass degrees of freedom that the interacting sector in the IR CFT on a single monopole string which does not degenerate to monopole strings in local SCFTs is unitary and  has normalizable vacua.
Thus the small $\mathcal{N}=(4,4)$ superconformal algebra cannot be realized in the non-trivial CFTs on monopole strings unless there exists an accidental $SU(2)$ symmetry in IR\footnote{If a 5d monopole string comes from a 6d self-dual string on $S^1$ without twist, the worldsheet theory can flow to a non-trivial SCFT with  $(4,4)$ SUSY enhancement.
This is because in this case the symmetry $SO(3)_R$ for the transverse $\mathbb{R}^3$ rotation enhances to $SU(2)_l\times SU(2)_r$ in IR, and the $SU(2)_l$ and $SU(2)_r$ can become the left-moving and the right-moving R-symmetries, respectively, of the $(4,4)$ superconformal algebra. We expect that the 6d supergravity strings studied in \cite{Kim:2019vuc} with $Q\cdot a=0$ have a $\mathcal{N}=(4,4)$ SUSY enhancement on a non-trivial interacting sector. When geometrically realized, such 6d strings have the same number of left-moving bosons and fermions, $N_L^B=N_L^F=4(g+1)$ when the pull-back $\widehat{C}$ wrapped by dual M5 brane is a trivial fibration $C \times E$, where $E$ is an elliptic curve and $g$ is the genus of the curve of the string in the base. This is consistent with $(4,4)$ SUSY (or (8,8) SUSY when $g=1$) enhancement.}.

This argument however cannot rule out the possibility of $\mathcal{N}=(4,4)$ SUSY enhancement when the interacting sector in the IR CFT is trivial. It is possible that the IR worldsheet CFT consists of only the center-of-mass degrees of freedom so that the interacting CFT sector is trivial. The above argument does not hold for the center-of-mass sector due to the non-compact free bosons parametrizing the transverse motion of the string. The worldsheet theory can flow in the infrared to a free theory consisting of the $\mathcal{N}=(4,4)$ center-of-mass multiplet formed by 3 non-compact bosons $X_{\mu=1,2,3}$ and a compact scalar $\phi$ and 4 chiral and anti-chiral fermions $\lambda_\pm^\alpha$ where $\alpha$ is the doublet index of $SU(2)_R$. In this case, the IR R-symmetry is identified with the $SU(2)_R$ symmetry, and the central charges are $c_L=c_R=6$. This implies that the $\mathcal{N}=(4,4)$ SUSY enhancement can occur only if $c_R-c_L=0$ and $k_R=0$, therefore only if $C_{IJK}q^Iq^Jq^K=C_Iq^I=0$.

The worldsheet theory can have a further enhancement to $\mathcal{N}=(8,8)$ SUSY. In this case, the worldsheet theory consists of a free $(8,8)$ center-of-mass multiplet and the interacting sector in the IR CFT is again trivial. The central charges from the free $(8,8)$ multiplet are $c_L=c_R=12$. Thus, this string has $C_{IJK}q^Iq^Jq^K=C_Iq^I=0$. This string lives in the 5d supergravity theory with $32$ supercharges. This string amounts to a M5-brane wrapping an Abelian surface with irregularity $q=2$ in M-theory compactification.

Lastly, the worldsheet SUSY can enhance to $\mathcal{N}=(0,8)$ supersymmetry. The strings coupled to 5d bulk gravity with such enhancement are those in the 5d supergravity theories with 16 supercharges. It was conjectured in \cite{Kim:2019ths} that such strings have central charges $c_L=24$ and $c_R=12$ coming from only the $(0,8)$ center-of-mass modes. This indicates that the $\mathcal{N}=(0,8)$ enhancement can occur in the worldsheet theory only when $C_{IJK}q^Iq^Jq^K=0$ and $C_Iq^I=24$.
An M5-brane wrapping a K3 surface of Table \ref{tb:minimal-S} leads to such a monopole string with $(0,8)$ supersymmetry.

\paragraph{Central charges}
In this paper we will focus on magnetic monopole strings whose worldsheet CFTs exhibit no symmetry enhancement. Such strings have the $SU(2)_R$ as the R-symmetry of $(0,4)$ superconformal algebra in the IR CFT. The definition and some important properties of these special strings will be introduced in the following subsection.

For such a string, the precise central charges can be computed by combining anomaly coefficients in (\ref{eq:anomaly}). The worldsheet theory for string charge $q^I$ has the central charges as
\begin{equation}\label{eq:central}
	c_L = C_{IJK}q^Iq^Jq^K + C_Iq^I \ , \quad c_R = C_{IJK}q^Iq^Jq^K + \frac{1}{2} C_Iq^I \ .
\end{equation}
So the central charges are fully determined by the Chern-Simons levels $C_{IJK}$ and $C_I$ as well as the string charge $q^I$.
We remark that since the 't Hooft anomaly coefficient $k_R$ for the $SU(2)_R$ is quantized to be an integer, the right-moving central charge is also quantized as $c_R \in 6\mathbb{Z}$. As we will see later, this provides a strong constraint on the Chern-Simons levels in the effective supergravity action.

These central charges involve the contributions from the center-of-mass degrees of freedom. The center-of-mass modes consist of four bosons $(X_{\alpha\beta}^+,\phi)$ and four fermions $\lambda_+^\alpha$ in the right-moving sector and three bosons $X_{\alpha\beta}^-$ in the left-moving sector \cite{Mizoguchi:1998wv}. They form a free hypermultiplet of $\mathcal{N}=(0,4)$ supersymmetry. Their contribution to the central charges can be easily read off from the free field content as
\begin{equation}\label{eq:central-com}
	c^{\rm com}_L = 3 \ , \quad c^{\rm com}_R = 6 \ .
\end{equation}
The center-of-mass modes decouple from the interacting CFT in IR. 

Therefore, the central charges ($\hat{c}_L,\hat{c}_R$) of the interacting SCFT on a single string with magnetic charge $q^I$ are given by
\begin{eqnarray}\label{eq:central-c2}
	\hat{c}_L &\equiv& c_L - c^{\rm com}_L = C_{IJK}q^Iq^Jq^K + C_Iq^I -3 \ , \nonumber \\
	\hat{c}_R &\equiv& c_R - c^{\rm com}_R =  C_{IJK}q^Iq^Jq^K + \frac{1}{2}C_Iq^I - 6 \ .
\end{eqnarray}

The worldsheet theory can carry the current algebras for the bulk gauge symmetry. The 't Hooft anomaly $k_{IJ}$ of the current algebra can also be extracted from the anomaly polynomial as
\begin{equation}\label{eq:abelian-level}
	k_{IJ} = C_{IJK}q^K \ ,
\end{equation}
for the mixed anomaly between two Abelian currents $J^I$ and $J^J$. In our convension, the right- (or left-) moving charged fields add positive (or negative) contributions to the anomaly coefficient $k_{IJ}$.

As we discussed the bulk Abelian gauge symmetry can enhance to non-Abelian symmetry, say $G_i$, at some special points of the Coulomb branch. In this case, the string worldsheet theory can furnish a representation of the current algebra for the non-Abelian symmetry. The chiral fields realizing the current algebra yield 't Hooft anomaly, which can be read off from (\ref{eq:anomaly}), of the form
\begin{equation}\label{eq:non-abelian-level}
	-\frac{1}{4}k_i {\rm Tr}F_i^2 \ \quad {\rm with} \quad  k_i = -h_{i,\alpha}q^\alpha \ ,
\end{equation}
where  $F_i$ is the field strength of $G_i$ and $h_{i,\alpha}$ is the coefficient in the gauge coupling $h_i$ for $G_i$ in the bulk effective action.
The 't Hooft anomaly coefficient $k_i$ for the non-Abelian symmetry is quantized as an integer number.
The $k_i$ is related to the level for the current algebra of the symmetry $G_i$. The level $k$ current algebra of $G_i$ realized by right-movers (or left-movers) provides $+k$ (or $-k$) contribution to the anomaly coefficient $k_i$.

As a simple example, let us consider the M-theory compactification on the quintic Calabi-Yau 3-fold discussed in the next section. This engineers a 5d $\mathcal{N}=1$ supergravity theory with a single $U(1)$ gauge symmetry at low energy.
The effective action is characterized by the cubic and the linear Chern-Simons levels given by
\begin{equation}
	C_{000} = 5 \quad {\rm and} \quad C_0 = 50 \ .
\end{equation}
Now consider a monopole string with positive magnetic charge $q$ for the $U(1)$ gauge symmetry. Using (\ref{eq:central}) and (\ref{eq:abelian-level}), one can easily compute the central charges of the 2d CFT on the string,
\begin{equation}
	c_L = 5q^3 + 50q \ , \quad c_R = 5q^3 + 25q \ ,
\end{equation}
and the 't Hooft anomaly of the $U(1)$ current,
\begin{equation}
	k_{U(1)} = 5q \ .
\end{equation}
This implies that the worldsheet theory has a $U(1)$ current algebra with level $5q$ in the right-moving sector.

\subsection{Supergravity strings}\label{subsec:sugra}
We will now introduce a special class of monopole strings called {\it supergravity strings}. The supergravity strings are  1/2 BPS objects that  appear only in  gravity theories and  not in local CFTs. In this section, we will make this distinction clear and investigate consistency conditions on supergravity strings together with their implications for 5d supergravity theories.
If the supergravity is geometrically engineered then these strings can be understood as M5 branes wrapping {\it semi-ample} 4-cycles in the geometry. In particular, the distinction between supergravity strings and local strings can be understood through the different properties that the 4-cycles need to satisfy. We will also investigate these properties in detail which will lead us to a geometric definition of supergravity strings studied in more detail in Section \ref{sec:geosugra}.

The BPS states of the 5d supergravity are electrically charged particles and the dual magnetically charged monopole strings. All BPS states are expected to have non-negative masses and non-negative tensions on the Coulomb branch, which essentially defines the Coulomb branch of the scalar vevs in the vector multiplets. We first propose that the Coulomb branch $\mathcal{C}$ is the space of the scalar moduli $\phi^I$ bounded by the set of hyperplanes where some BPS particle states become massless:
\begin{equation}
	\mathcal{C} = \{\phi^I\,,I=1,\cdots,\mathbf{r}\,|\,m^2(\phi^I)\ge0\} \ .
\end{equation}
Here $m^2(\phi^I)\ge0$ denotes that mass squared of all BPS particles are non-negative at the point labelled by $\phi^I$. We also conjecture that if all BPS particles have non-negative mass squared, then the monopole string tensions are also non-negative at the point. This follows from the fact that if the volumes of all 2-cycles are non-negative, the volumes of 4-cycles are also non-negative in Calabi-Yau threefolds.

\begin{framed}
\noindent {\bf Definition:} A 1/2 BPS magnetic monopole string on the Coulomb branch $\mathcal{C}$ in a 5d supergravity theory defines a {\it supergravity string} if all supersymmetrically compatible BPS particle states in the theory carry non-negative electric charge $e(A_\mu)$ under the dual Abelian gauge field $A_\mu$:
\begin{equation}\label{eq:supergravity-strings}
	e_i(A_\mu) \ge 0 \ ,
\end{equation}
where $i$ runs over all particles obeying the BPS mass formula $|m|=\sum_Ie_I \phi^I$.
\end{framed}

The dual gauge field $A_\mu$ in this definition precisely means the Abelian gauge field whose positive minimal magnetic charge is carried by the supergravity string with tension $T=\partial \mathcal{F}>0$. We remark that the supergravity string here is defined with respect to a particular pair of supercharges shared by the supergravity string and the BPS particles taken into account in (\ref{eq:supergravity-strings}).
For a given supergravity string preserving these supercharges, the electric charge condition (\ref{eq:supergravity-strings}) holds for all BPS particles preserving the same supercharges, but does not need to hold for anti-BPS particles satisfying $|m|=-\sum_Ie_I\phi^I$ that preserve another set of supercharges. This distinction between BPS and anti-BPS states allows one to clearly distinguish the supergravity string from other strings. In the followings, BPS states we will use refer to the states preserving this pair of supercharges.

We now claim that the supergravity strings exist only in supergravity theories, while local theories such as 5d SCFTs cannot have any supergravity strings. This property will enable us to explore some distinguished features of gravitational theories by using the supergravity strings.
We note that the BPS W-boson of $U(1)\subset SU(2)$ gauge symmetry in a local gauge theory, called $E_1$ theory, in this convention has negative charge $-2$ under the $U(1)$ symmetry. Similarly, on the Coulomb branch the BPS W-bosons of a non-Abelian gauge group $G$ carry electric gauge charges $e_a=-(\mathcal{K}_{ab})$ of the $U(1)^r\subset G$ gauge fields $A^b_\mu$, where $\mathcal{K}_{ab}$ is the Cartan matrix of $G$ with rank $r$. This implies that in local 5d gauge theories, the BPS monopole strings with magnetic charge $q^a>0$ of the dual gauge fields $A^a_\mu$ cannot satisfy (\ref{eq:supergravity-strings}) and thus cannot be supergravity strings. More generally, we expect that supergravity strings cannot reside in any local theories. This can be proven for the cases admitting geometric constructions using geometric properties of special 4-cycles related to supergravity strings as we will see in the next section.

\begin{framed}
\noindent {\bf Conjectures}
	\begin{enumerate}
		\item Supergravity strings exist only in supergravity theories.
		\item The worldsheet theory on a supergravity string with magnetic charge $q^I$ flows to a $(0,4)$ SCFT with $SU(2)_R$ R-symmetry if $C_{IJK}q^Iq^Jq^K>0$.
	\end{enumerate}
\end{framed}

The condition $C_{IJK}q^Iq^Jq^K>0$ for the second conjecture rules out emergence of accidental symmetries in the worldsheet theories. For the strings with this condition, the R-symmetry in the infrared worldsheet CFT will be the $SU(2)_R$ symmetry. We are interested in only this type of supergravity strings in this paper.

As presented above, on the other hand, it is possible that the worldsheet theories on monopole strings with $C_{IJK}q^Iq^Jq^K=0$ have the SUSY enhancements or come from the strings in the 6d supergravity theories.  The 6d supergravity strings were already studied in \cite{Kim:2019vuc} by employing the same idea we use in this paper. The monopole strings hosting interacting CFTs with $\mathcal{N}=(4,4)$ SUSY are also a part of the 6d supergravity strings. The supergravity strings with $\mathcal{N}=(0,8)$ SUSY were studied in \cite{Kim:2019ths} and the strings with $\mathcal{N}=(8,8)$ SUSY come from Type II strings compactified on $T^5$.

Supergravity strings are magnetic sources for the gauge fields in a supergravity theory and they should exist due to the completeness assumption for the spectrum in the gravity theory \cite{Polchinski:2003bq,Banks:2010zn}.   We will be assuming a stronger version of completeness assumption because we will in addition assume that the corresponding states in the spectrum are represented by BPS objects.  This is obvious in geometry because any semi-ample divisor $\mathcal{D}$ with $\mathcal{D}^3>0$, which is defined in the next section, is effective and thus can be wrapped by an M5-brane. The wrapped M5-brane always leads to a supergravity string in the 5d gravity theory.

We expect that the worldsheet degrees of freedom living on a supergravity string satisfy unitary conditions. Basic unitary conditions are the following:
\begin{enumerate}
	\item The central charges of the interacting 2d SCFT on a supergravity string with charge $q^I$ are given by (\ref{eq:central-c2}) and they have to be non-negative
	\begin{equation}
	\hat{c}_L \ge 0 \ , \quad \hat{c}_R \ge 0 \ .
	\end{equation}
	\item The right-moving central charge is quantized as
	\begin{equation}
		c_R = C_{IJK}q^Iq^Jq^K+\frac{1}{2}C_Iq^I \in 6\mathbb{Z} \ .
	\end{equation}
	\item The tension of the supergravity string eq. (\ref{eq:Tension}) is always non-negative, i.e. $T_I\geq 0$, on the Coulomb branch $\mathcal{C}$.
\end{enumerate}

From the condition (\ref{eq:supergravity-strings}) for supergravity strings, one can find an interesting relation between supergravity strings and Coulomb branch in supergravity theories. The condition (\ref{eq:supergravity-strings}) tells us that the positive scalar vev in the vector multiplet\footnote{This scalar vev can also be in a hypermultiplet if the dual gauge field is the graviphoton.} dual to a supergravity string by itself without turning on other scalar vevs parametrizes a direction of moduli space where all particle states have non-negative mass squared. Therefore, this positive scalar vev necessarily lies within the Coulomb branch in the supergravity theory. Conversely, one may be able to find associated supergravity strings for any line on the Coulomb branch parametrized by a positive real number.

\subsection{General supergravity conditions}\label{sec:sugracon}
In this subsection, we will discuss unitarity conditions on supergravity strings without  assuming geometric embeddings of the gravity theory.

We will also argue that some geometric conditions discussed in the next section can be interpreted as physical constraints on supergravity strings.

Consider a BPS monopole string with magnetic charge $q^I$ in the 5d gravity theory and suppose that this string is a supergravity string. Then the central charges $\hat{c}_R$ and $\hat{c}_L$ in (\ref{eq:central-c2}) for the 2d IR CFT should be non-negative. This is an obvious condition for unitarity of the supergravity string.

More conditions can be found by using the 't Hooft anomalies given in (\ref{eq:abelian-level}) and in (\ref{eq:non-abelian-level}) for the Abelian and the non-Abelian symmetries, respectively.
For this, let us first clarify the relation between the 't Hooft anomaly coefficients and the levels of Kac-Moody current algebras of $G$ in the 2d CFTs. 
For a non-Abelian group $G_i$, the 't Hooft anomaly coefficient $k_i$ given in (\ref{eq:non-abelian-level}) is required to be an integer by quantization condition. The 't Hooft anomaly receives positive contributions from right-moving modes while receives negative contributions from left-moving modes.
Thus the level $k$ current algebra adds $+k$ or $-k$ to the associated anomaly coefficient according to its chirality.
It then follows that a non-zero coefficient $k_i$ implies that the 2d CFT contains at least a Kac-Moody current algebra for $G_i$ with level $k= |k_i|$ in the right-moving sector when $k_i>0$ or in the left-moving sector when $k_i<0$.

The 't Hooft anomalies for Abelian groups can mix each other. The anomaly coefficient $k_{IJ}$ in (\ref{eq:abelian-level}) is a symmetric matrix and the eigenvalues are identified with the levels of Abelian current algebras. The precise values of the levels are not important in our discussions. However, the signature of the anomaly coefficient $k_{IJ}$ is of some significance. It encodes the lower bounds on the number of representations of Abelian current algebras in the left- or right-moving sector. If the anomaly coefficient $k_{IJ}$ has signature $({n_+},{n_-})$, where $n_+$ and $n_-$ denote the number of positive and negative eigenvalues respectively, the worldsheet CFT necessarily involves at least $n_+$ Abelian current algebras in the right-moving sector and $n_-$ Abelian current algebras in the left-moving sector.

We shall now argue that the signatures of 't Hooft anomaly coefficients in the 2d CFT on a supergravity string are restricted.
The current algebras of a worldsheet CFT are realized by zero modes of the bulk fields on the magnetic monopole string background. 
The right-movers come from the Goldstone modes of broken symmetry generators including three position zero modes and a compact bosonic mode for the broken $U(1)$ gauge symmetry as well as their fermionic partners.
On the other hand the other charged matter fields coupled to the supergravity string leave only fermionic zero modes that are left-movers on the worldsheet \cite{Boyarsky:2002ck}. This means that the current algebras of gauge groups, but a single Abelian group, necessarily sit in the left-moving sector.

In the next section we define supergravity strings geometrically as M5-brane states wrapping semi-ample divisors  by definition \ref{eq:condition-semi-ample}. Note that a semi-ample divisor has a non-negative intersection with every divisor in the threefold. This implies that a semi-ample divisor at the intersection with the gauge divisors for gauge group $G$ can provide only hypermultiplets charge under the gauge symmetry. These hypermultiplets give fermionic zero modes in the left-moving sector on the string background.

We can rephrase this as the following condition on the signatures of the 't Hooft anomaly coefficients of the gauge groups under which the string is charged:
\begin{equation}\label{eq:signature}
{\rm sig}(k_{IJ}) = (1,r-1) \ ,
\end{equation}
for Abelian groups where $r$ is the rank of Abelian groups, and
\begin{equation}\label{eq:level-G}
k_i < 0 \ ,
\end{equation}
for all non-Abelian groups.
This condition on the 't Hooft anomalies and thus on the levels of current algebras is one of the special features of supergravity strings. Moreover, this provides a field theory interpretation of the Hodge index theorem \ref{HI}  for the signature of the intersection pairing of semi-ample divisors in geometry.

Note that one of ${\rm sig}(k_{IJ})$ for a supergravity string is always positive. This positive level is the center-of-mass contribution. The corresponding $U(1)$ current algebra is generated by the compact right-moving scalar field $\phi$ in the $\mathcal{N}=(0,4)$ center-of-mass free hypermultiplet. In the geometric setting, this current algebra is generated by the divisor class itself for the monopole string \cite{Maldacena:1997de}. After subtracting this center-of-mass contribution, all other levels in the interacting worldsheet CFT are negative meaning that the current algebras of gauge symmetries, but that of the $SU(2)_R$ symmetry, are realized in the left-moving sector.

We remark here that the supergravity strings in the 6d supergravity theories share the same property. As studied in \cite{Kim:2019vuc}, the worldsheet CFTs on 6d supergravity strings contain only left-moving current algebras. This property was used to distinguish supergravity strings from the instanton strings in 6d local SCFTs or little string theories.

A unitary realization of a current algebra contributes to the central charges in the 2d CFT. The central charge contribution from an Abelian current algebra is $c_{U(1)}=1$. For a non-Abelian current algebra of $G_i$ at level $k$, the central charge contribution is
\begin{equation}
c_{G_i} = \frac{k\cdot {\rm dim}G_i}{k+h^\vee_i} \ ,
\end{equation}
where ${\rm dim}G_i$ is the dimension and $h^\vee_i$ is the dual Coxeter number of group $G_i$. Unitarity requires the level $k$ to be positive definite.
As discussed, all the current algebras are sitting in the left-moving sector in the interacting CFTs on supergravity strings. 
From this we find an inequality between the levels and the left-moving central charge
\begin{equation}\label{eq:level-c}
n + \sum_i c_{G_i} \le \hat{c}_L \ ,
\end{equation} 
where $n$ is the number of Abelian gauge groups and $G_i$'s are the non-Abelian groups whose current algebras are realized in the interacting CFT. 

The unitary 2d $(0,4)$ CFT on a supergravity string must satisfy this inequality together with the conditions $\hat{c}_L,\hat{c}_R\ge0$. If these conditions are violated by a supergravity string, then it means the string cannot host a unitary CFT that cancels the anomaly inflow arising from the bulk 5d supergravity theory.
As a result the supergravity string cannot consistently couple to the 5d supergravity. By the completeness of string spectrum, the supergravity theory is therefore in the Swampland.

One can find more conditions on the supergravity strings from the properties of black holes. For example, let us consider a cone of monopole strings given by a linear combination of supergravity strings as $\mathcal{D}=\sum_I n_Iq^I$ with positive coefficients $n_I$. This defines a K\"ahler cone of the low-energy theory. Choose now an arbitrary string of $D$ inside the K\"ahler cone with positive coefficients $n_I>0$. Then the large multiple of the chosen string, i.e. $mD$ with $m\gg1$, is expected to form a black string state with the entropy $S\propto \sqrt{\hat{c}_L}$. At large $m$, the cubic terms in the $\hat{c}_L$ scale as $m^3$ and will dominate the other terms. The central charge of the black string should be positive by unitarity. This immediately restricts the cubic term for any supergravity string to be positive semi-definite,
\begin{equation}\label{pos}
C_{IJK}q^Iq^Jq^K \ge 0 \, .
\end{equation}
In geometry such string $mD$ at large $m$ amounts to a very ample divisor.

 Indeed, the wrapped M5-brane on a very ample divisor with large $c_L$ form a black string \cite{Maldacena:1997de}.

\section{M-theory on Calabi-Yau threefolds}\label{sec:Mtheory}
In the previous section we studied general aspects of 5d $\mathcal{N}=1$ supergravity theories.
However, a  large class of such supergravity theories can be engineered by compactification of M-theory on compact Calabi-Yau threefolds (CY$_3$'s) with $SU(3)$ holonomy.
In M-theory compactification, the Coulomb branch of the moduli space in the 5d theory is identified with the K\"ahler moduli space of the CY 3-fold. The K\"ahler moduli space is parametrized by K\"ahler parameters $\phi^I$ associated to an integral basis $\omega_I$ for $H^{1,1}(X)$ in a threefold $X$.
The action and the matter content in the effective five-dimensional theory are specified by topological data of the K\"ahler moduli space.

Let us expand the three-form potential $C_3$ in 11d supergravity in terms of the basis two-form classes $\omega_I$ of $X$ as
\begin{equation}
C_3 = A^I\wedge \omega_I \ ,
\end{equation}
where $A^I=A_\mu^I dx^\mu$ with $I=0,\cdots,h^{1,1}(X)-1$ are the 1-form vector fields along the non-compact 5d spacetime. A particular linear combination of the 1-form fields will become the graviphoton field in the gravity multiplet and the remaining $n_V=h^{1,1}(X)-1$ vector fields will become the $U(1)$ gauge fields in the vector multiplets in the low-energy supergravity theory. The 11d supergravity action integrated on the threefold $X$ reduces to the effective action of the 5d supergravity up to four-derivative terms written in terms of massless supermultiplets.

The reduction of the 11d Chern-Simons term on $X$ leads to the 5d cubic Chern-Simons term \cite{Papadopoulos:1995da}:
\begin{equation}
S_{CS}=-\frac{1}{6}\int_{\mathcal{M}_{5}\times X}C_3 \wedge G_4\wedge G_4 = -\frac{1}{6}C_{IJK}\int_{\mathcal{M}_5}A^I \wedge F^J \wedge F^J \ ,
\end{equation}
where $G_4=dC_3$ and $F^I=dA^I$. In this expression, the triple intersection number
\begin{equation}
C_{IJK} \equiv \int_X \omega_I\wedge \omega_J \wedge \omega_K \ ,
\end{equation}
counts the intersection numbers of 4-cycles dual to $\omega_I$ in $X$. This triple intersection numbers naturally reduce to the cubic Chern-Simons coefficients $C_{IJK}$ in the 5d effective action.

We can now write the K\"ahler form $J$ in this basis as
\begin{equation}
J = \sum_I \phi^I \omega_I \ , \quad I=0,1,\cdots,h^{1,1}(X)-1 \ ,
\end{equation}
where $\phi^I$ are the $h^{1,1}(X)=n_v+1$ K\"ahler moduli. Note that one of these moduli controlling the overall volume of  $X$ becomes a scalar component in a hypermultiplet and the other moduli are mapped to the scalar expectation values in the vector multiplets in the low-energy theory.
The total volume of $X$ measured with respect to $J$ is then given by
\begin{equation}
\mathcal{F} = \frac{1}{6}\int_X J\wedge J \wedge J = \frac{1}{6}C_{IJK}\phi^I\phi^J\phi^K \ .
\end{equation}

We shall fix the value of this volume (so a hypermultiplet scalar) by a constraint $\mathcal{F}=1$. The remaining scalars $\phi^I$ subject to this constraint form an $n_V$ dimensional K\"ahler moduli space in $X$. This  K\"ahler moduli space is identified with the Coulomb branch of the moduli space in the 5d supergravity theory. 

The metric on the K\"ahler moduli space is geometrically defined as
\begin{equation}
G_{IJ} = \frac{1}{2}\int_X \omega_I\wedge *\omega_J = -\frac{1}{2}\partial_I\partial_J(\log \mathcal{F})|_{\mathcal{F}=1} \ ,
\end{equation}
where the $*$ denotes the Hodge dual taken in the  internal Calabi-Yau manifold. This metric agrees with that of the supergravity theory given in (\ref{eq:metric}) when we identify the volume $\mathcal{F}$ of $X$ with the prepotential in the 5d gravity theory.

The spectrum of charged objects under the 5d gauge symmetry originate from  M2/M5-branes in M-theory wrapped on 2/4-cycles in the 3-fold. The M2-brane has a three-dimensional worldvolume carrying unit electric charge of the 3-form potential $C_3$. The worldvolume of  M2-branes can wrap on holomorphic (and also effective) 2-cycles in the internal 3-fold.  The wrapped M2-branes give rise to electrically charged BPS particles coupled to the vector fields $A^I$ for the 2-cycles in the 5d theory. In this case, the mass of the BPS particle is proportional to the volume of the 2-cycle. So the BPS particles coming from the wrapped M2-branes are massive at generic points on the K\"ahler moduli where all 2- and 4-cycles have finite volume. Thus they can be integrated out and do not appear in the spectrum of the low-energy effective theory.

At certain special values of the K\"ahler moduli $\phi^I$, some 2-cycles (and also 4-cycles) shrink to zero size and the 3-fold $X$ becomes singular. The singularity with vanishing cycles can support a non-Abelian gauge algebra $G$ and the M2-branes wrapping shrinking 2-cycles give rise to massless charged states in the 5d field theory. In particular, the massless vector states can participate in the gauge symmetry enhancement to the non-Abelian symmetry $G$ supported along the singularity of the 3-fold.

The M5-brane is a magnetically charged object with respect to the 3-form potential.
The six-dimensional worldvolume of the M5-brane can wrap around holomorphic 4-cycles labelled by $\omega_I$ in the internal threefold. The remaining two-dimensional worldvolume stretches along the 5d non-compact spacetime. Thus the wrapped M5-brane yields a magnetically charged monopole string of the Abelian gauge field $A^I_\mu$ in the 5d supergravity theory.  

The tension of the BPS monopole string is set by the volumes of 4-cycles in the Calabi-Yau manifold. For the string with unit magnetic charge of $A^I$, the volume of a basic 4-cycle $\omega_I$, so the string tension, is given by
\begin{eqnarray}
T_I &=& \partial_I \mathcal{F} = \frac{1}{2}\int_X \omega_I\wedge J \wedge J \nonumber \\
&=& \frac{1}{2}C_{IJK}\phi^J\phi^K \ .
\end{eqnarray}
The string tension is always positive within the K\"ahler cone of a smooth 3-fold $X$.

In the singular limit of $X$, as stated above, some 4-cycles can collapse to a point or to a collection of 2-cycles. Then the M5-branes wrapping the set of collapsing 4-cycles become tensionless strings.
The low-energy theory in the neighborhood of the singular locus when gravity is decoupled reduces to a local 5d SCFT strongly interacting with the tensionless strings \cite{Seiberg:1996bd,Morrison:1996xf,Douglas:1996xp}. However, we are not interested such tensionless strings in local 5d SCFTs.
In the following sections, we will focus only on the wrapped M5-brane states over 4-cycles which never collapse to zero size in the K\"ahler moduli space of the Calabi-Yau threefold. These string states are related to the supergravity strings as we will discuss now.

\subsection{Supergravity strings from wrapped M5 branes}\label{sec:geosugra}
In Section \ref{subsec:sugra} we defined a certain class of 5d monopole strings called \textit{supergravity strings} which appear only in supergravity theories.
In fact the definition of supergravity strings was motivated by geometric considerations of monopole strings and associated 4-cycles in Calabi-Yau geometry.
In the case that the supegravity is geometrically engineered through M-theory on a Calabi-Yau threefold, the BPS states can be understood as M2- and M5-branes wrapping holomorphic 2- and 4-cycles respectively. 
Therefore,  the 5d monopole strings arise from  M5-branes wrapping 4-cycles represented by  effective divisor classes $	[\mathcal{D} ]\in Pic(X)\cong H^2(X,\IZ)$ of the Calabi-Yau 3-fold $X$. The divisor class being effective means that it can be represented by an effective divisor $\mathcal{D}$, i.e.\ $\mathcal{D}$ is a non-negative linear combination of surfaces $D_i$ (possibly singular): $\mathcal{D}=\sum_i n_iD_i$ (Definition~\ref{effective2}).

In order for a monopole string to be a \textit{supergravity string} as defined in (\ref{eq:supergravity-strings}), we require the electric charges of all BPS particles for the dual gauge field to be non-negative.  A BPS particle is the M2-brane wrapping a curve $C$ and its electric charge for the gauge field dual to a divisor class $\mathcal{D}$ is given by the intersection of $C$ and $\mathcal{D}$.
We thus claim that

\begin{framed}
\noindent
The monopole string wrapping an effective divisor $\mathcal{D}$ is a \textit{supergravity string} precisely when 
\begin{equation}\label{eq:condition-semi-ample}
\mathcal{D}\cdot C\ge0 \ \text{\ for all curves } \ C\subset X.
\end{equation}
\end{framed}

The above condition on a divisor $\mathcal{D}$  is called \emph{nef} in the algebraic geometry literature (Definition~\ref{nef}).  So our supergravity strings arise from M5-branes wrapping nef and effective divisors.  Assuming a conjecture which we will formulate and motivate below, this is equivalent to just nef, or just semi-ample (Definition~\ref{semidef}).

The nef condition is closely related to the condition of a divisor being ample, or equivalently that its cohomology class is a K\"ahler class.
Like nef divisors, ample divisors are characterized by its intersections, but the condition is more complicated: for $\mathcal{D}$ to be ample we require $\mathcal{D}^3>0$, $\mathcal{D}^2\cdot S>0$ for all surfaces $S\subset X$, and $\mathcal{D} \cdot C >0$ for all curves $C\subset X$.  This is the Nakai-Moishezon criterion for ampleness (Theorem~\ref{semi-ample}).
In particular, from the point of view of  divisors, the K\"ahler cone $ \mathcal{K}(X)$ is generated by ample divisor classes.

As explained immediately after Theorem~\ref{Nef}, the closure of the K\"ahler cone $\mathcal{\overline{K}}(X)$  is simply the \emph{nef cone} spanned by the classes of nef divisors. 
Therefore, the BPS states can be understood as M2-branes wrapping curves in the Mori cone $M(X)$ and the supergravity strings are the dual M5-branes wrapping surfaces in the dual cone $\mathcal{\overline{K}}(X)$.

The effectiveness of the divisor class that the M5-brane wraps is required in order for it  to be represented by a surface.
However, the basic criterion of distinguishing supergravity strings from other monopole strings wrapping surfaces is the condition that the 4-cycle is nef.

\begin{framed}
\noindent
\textbf{Conjecture: }\textit{any nef divisor is linearly equivalent to an effective divisor, and so can be represented by a surface}.  
\end{framed}

Assuming the conjecture, we only need the nef condition in order to get a supergravity string in the geometric setting.  This conjecture is an open question of mathematics which has been discussed in the mathematics literature for several decades, e.g.~\cite{Lazarsfeld:2004pag,OguisoK}.

However, we are making this conjecture based on considerations of physics, specifically the completeness of spectrum hypothesis.  But first, we explain a bit more of the mathematical background, referring to the appendices for more complete definitions, proofs, and references.  Assuming that the divisor $\mathcal{D}$ is nef, we have the following:
\begin{itemize}
	\item If $\mathcal{D}^3>0$ we know that $\mathcal{D}$ is in fact big (Definition~\ref{def:big})  by  Theorem~\ref{prop:big}, which implies that is linearly equivalent to an effective divisor (Corollary~\ref{cor:eff}).
	\item  If $\mathcal{D}^3=0$ but $\mathcal{D}^2\neq0$,  and $c_2(X)\cdot \mathcal{D}>0$ then by \cite{OguisoK} we know that the divisor is also linearly equivalent to an effective divisor. 
	\item If we only require $c_2(X)\cdot \mathcal{D}>0$ then by \cite{OguisoK} we know that there is a multiple of the divisor $m\mathcal{D}$ that is linearly equivalent to an effective divisor for $m \gg 1$. 
\end{itemize}
Mathematically, there is no known proof that we can take $m=1$ in the last case. However, since $m\mathcal{D}$ is an effective nef divisor then an M5 brane wrapping it gives rise to a supergravity string.  In addition, by the completeness of spectrum hypothesis  we know that the charge lattice should include all minimal charge states for a given state. Therefore this supports the claim that if $m\mathcal{D}$ gives rise to a supergravity string then the class of $\mathcal{D}$ should too.
This implies that $\mathcal{D}$ should be  effective as well in order to be represented by a physical surface.  

Moreover, if we consider the cases where $\mathcal{D}^3=c_2\cdot \mathcal{D} =0$, by Section \ref{sec:Monopole} we expect  that those cases are presenting supersymmetry enhancement of the monopole string worldsheet. In particular, these divisors lead to (4,4) and (8,8) supersymmetry enhancement.  As we will see in the next section, (hyper-)elliptic and abelian surfaces of this type can be understood as giving rise to exactly that amount of supersymmetry respectively. Therefore, we expect that the divisors associated to $\mathcal{D}^3=c_2\cdot \mathcal{D} =0$ are also effective.

(Semi-)Ample divisors in a CY$_3$ have several distinguished features.
Let us consider a subspace of the K\"ahler moduli space parametrized solely by the K\"ahler parameter for a given ample divisor $\mathcal{D}$. The volume of a 2-cycle $C$ in $X$ is determined with respect to the K\"ahler form $J$ as ${\rm vol}(C)=J\cdot C$. On this subspace, the K\"ahler form can be written as $J=\phi \mathcal{D}$ with the positive K\"ahler parameter $\phi$ for $\mathcal{D}$. Then the volume of a curve $C$ is given by $vol(C)=\phi\, \mathcal{D}\cdot C$.
Since the ample divisors positively intersect any $C$ in the Mori cone, the volume of every $C$ in the Mori cone is positive with the positive K\"ahler parameter $\phi$, i.e. $vol(C)>0$ for all curves $C\subset$ Mori cone. This implies that the 3-fold $X$ is smooth with positive volume on this subspace. Similarly, on a subspace of the Coulomb branch parametrized by a single positive K\"ahler parameter for a semi-ample divisor, the volume of every curve in the 3-fold is non-negative, thus $vol(C)\ge0$ for all curves $C\subset$ Mori cone.

\vspace{.5cm}
\noindent
\textbf{Remarks}
\begin{enumerate}
\item If $\mathcal{D}$ is nef and effective, then it is semi-ample (Definition~\ref{semidef}), which implies that the class of $m\mathcal{D}$ can be represented by a smooth surface for $m\gg1$ (see Bertini's Theorem~\ref{thm:bertini} and the remark immediately following).  With our main conjecture, we only need to assume $\mathcal{D}$ is nef.  
\item It can be seen using the results in Appendix~\ref{subsec:theorems} that the conjecture ``nef implies effective" is mathematically equivalent to the juxtaposition of the two conjectures ``nef implies semi-ample" and ``semi-ample implies effective".  If the conjecture ``nef implies effective" is true, then every nef divisor is nef and effective, hence semi-ample (Theorem~\ref{thm:sa}).  Furthermore, if a divisor is semi-ample, then it is nef (Proposition~\ref{prop:sanef}), hence semi-ample by the assumed conjecture.   The other direction of the equivalence is trivial.  In particular, our conjecture implies that \emph{the monopole strings which are supergravity strings are precisely the ones which wrap semi-ample divisor classes}. 
\item  The statement that all nef divisors are semi-ample was conjectured in \cite{OguisoK}.
\end{enumerate}

In the meantime we would like to distinguish the  supergravity strings from the monopole strings in local theories from a geometric viewpoint. Local field theories such as 5d SCFTs and 6d SCFTs on $S^1$ can be engineered by M-theory compactified on local non-compact Calabi-Yau 3-folds.  We illustrate this with local $\IP^2$.  Let $\mathcal{D}$ be $\IP^2$, thought of as a divisor in the local threefold.  Let $C$ be a curve of degree $d$ in this $\IP^2$.  Then $\mathcal{D}\cdot C=-3d$.  So $\mathcal{D}$ is not nef and the associated monopole string cannot be a supergravity string, even after embedding our local geometry in a compact geometry.   On the other hand, $(-\mathcal{D})\cdot C=3d>0$.  This is precisely the condition needed to get an SCFT!   Returning to mathematical terminology, $-\mathcal{D}$ is nef, if we understand nef on a noncompact threefold to be a condition on intersections with all compact curves.\footnote{See the remark following Theorem~\ref{Nef} for further discussion of this point.}  But $-\mathcal{D}$ is not effective.  Rather, it is \emph{anti-effective}, meaning that its negative is effective.  So on local $\IP^2$, we have an anti-effective nef divisor, which guarantees that we have an SCFT.\footnote{If we embed the divisor $\mathcal{D}=\IP^2$ inside a compact Calabi-Yau, then the anti-effective divisor $-\mathcal{D}$ cannot be nef.  To see this, take any curve $C$ which meets $\mathcal{D}$ in a finite nonzero number of points, for example the intersection of two general very ample divisors.  Then $(-\mathcal{D})\cdot C < 0$ and consequently $-\mathcal{D}$ is not nef.}

Recall that the geometries for local 5d (and also 6d) SCFTs are constructed by gluing ruled or rational surfaces and their blowups \cite{Jefferson:2018irk,Bhardwaj:2019fzv}. The monopole strings in these theories are constructed from M5 branes wrapping those surfaces, but they are not supergravity strings as the wrapped divisors are never nef.  By simply changing some signs in \cite{Jefferson:2018irk}, we summarize the local case by saying that we get a 5d or 6d SCFT when the local geometry supports an anti-effective nef divisor which includes each of the glued surfaces in its support.  Thus we claim that nef divisors exist only in compact CY 3-folds. This supports our claim that supergravity strings exist only in the 5d supergravity theories.

The supergravity strings are closely related to black strings (or MSW strings) studied in \cite{Maldacena:1997de,Minasian:1999qn}. One can find a black string solution with a smooth horizon when the corresponding M5-brane wraps on a {\it very ample} divisor with a large central charge  in the compact CY$_3$ \cite{Maldacena:1997de}. Theorem (\ref{Opt}) tells us that a very ample divisor can be constructed by considering a multiple of an ample divisor, i.e. 10$\mathcal{D}$ is very ample if $\mathcal{D}$ is ample. In the black string solution, the attractor mechanism forces the K\"ahler class at the horizon to be the divisor class wrapped by the M5-brane giving rise to the black string. Then the requirement for the volume of every 2-cycle near the horizon to be positive restricts divisors forming black strings to be ample. So the black string should come from a wrapped M5-brane over an ample divisor.
Furthermore, for the black hole solution being weakly curved, the ample divisor necessarily has large triple intersection number, implying that the divisor for the black string solution has to be very ample. This therefore means that a black string can always be written as a positive linear combination of K\"ahler cone $\mathcal{\overline{K}}(X)$ generators which are semi-ample divisors and thus is related to supergravity strings.

\subsection{Geometric conditions}\label{sec:geocon}
We can  extrapolate several geometric conditions by considering the various properties of  semi-ample, ample and  very ample divisors in a Calabi-Yau threefold.   Ideally, our supergravity strings would arise from wrapping smooth surfaces.  Any very ample divisor class has a smooth representative by Bertini's Theorem (Theorem~\ref{thm:bertini}), but this is not necessarily the case for semi-ample or even ample divisors.  An example of an ample divisor class that does not have a smooth representative is given in Appendix~\ref{app:singular}.  All we know is that some multiple of a semi-ample class has a smooth representative.  For ample divisor classes $\mathcal{D}$, we can bound the multiple needed as the class $5\mathcal{D}$ has a smooth representative as observed before.

If a semi-ample divisor class can be represented by a smooth surface $P$, we can say more. Using tools from algebraic geometry one can show that smooth semi-ample divisors are minimal surfaces\footnote{These surfaces are shown to be minimal in Appendix~\ref{subsec:theorems}, meaning that they are not isomorphic to the blow up of any other smooth surface.}
with Kodaira dimension $\kappa\geq 0$, as shown in Proposition~\ref{prop:kodairaprop}.   If $P$ is an ample divisor, or more generally if we have a divisor $P$ satisfying $P^3>0$ and $P\cdot C\ge 0$ for all curves $C$, then $b_1(P)=0$ and $\kappa=2$, also by Proposition~\ref{prop:kodairaprop}.

A complete classification of minimal projective surfaces is provided in Table \ref{tb:minimal-S} in terms of $\kappa$. 
The first row of  Table \ref{tb:minimal-S} represents surfaces with $\kappa =-\infty $ which can give monopole strings in local field theories,  not relevant for our current study. 
\begin{table}[h!]
	\centering
	\begin{tabular}{c|c|c}
		\hline
		$\kappa$ & K\"ahler surfaces $S$ & Type  \\
		\hline 
		$-\infty$ & $\mathbb{P}^2$, ruled surface $\mathbb{F}_n^g$  & Local Theories\\ 
		\hline
		$0$ & $K3$ & (0,8) susy enhancement. \\ 
		\hline
		$0$ & Enriques surface & 5d supergravity string\\ 
		\hline

		$0$ & hyperelliptic surface& (4,4) susy enhancement   \\ 
		\hline
	
		$0$ & abelian surface & (8,8) susy enhancement. \\ 
		\hline
		$1$ & minimal elliptic surface & 6d supergravity string on a circle\\ 
			 &  &  when it has a section (5d otherwise)\footnotemark.\\ 
		\hline
		$2$ & minimal surface of general type &  5d (0,4) supergravity string when  $b_1=0$ \\ 
		\hline
	\end{tabular}
	\caption{The first column is  the Kodaira dimension  $\kappa$ of the surface S. The second column presents the Enriques-Kodaira classification of minimal K\"ahler surfaces.   The third column describes the surfaces of the second column that could be 5d supergravity strings as defined in section \ref{subsec:sugra}.}
	\label{tb:minimal-S}
\end{table}
\footnotetext{Consider an elliptic threefold with a section and a curve C in base of the fibration wrapped by a D3 brane. The M/F-theory duality implies that the pull-back $\widehat{C}$ is wrapped by an M5 brane but such an elliptic surface always has a section. Therefore, an elliptic surface with no section does not correspond to a 6d string.}

For this reason, we only consider surfaces with $\kappa\geq 0 $ in the rest of this section.

\subsubsection{Computing the central charges $c_R,c_L$ from geomerty}\label{MSWcentral}
The central charges of the 2d SCFT on monopole strings in the 5d supergravity are related to  invariants of the associated surfaces. 
In particular, the degrees of freedom contributing to the central charges come  from the moduli of the  surface $P$, the two-form tensor fields and fermions on the worldvolume of the M5-branes on $P$. The computation of the central charges $c_R,c_L$, which we will review here, was done originally by \cite{Maldacena:1997de}, assuming very ampleness of the divisor $P$. However, the same arguments hold for an ample smooth divisor which we will assume for the next computation. Here $P$ is a  4-cycle and its cohomology class is  $[P]\in H^2(X,\IZ)$, which we will also  write as $P$ for simplicity.  In particular, $P$ can be expressed with respect to a basis $\omega_I$ of $H^2(X,\IZ) $ and charges $q^I\geq 0$ as $P=\sum q^I \omega_I$, meaning that the divisor is effective.

The left-moving central charge $c_L$ has no contributions coming from fermion zero modes since $b_1(P)=0$ by (\ref{Lht})\footnote{Assuming the Calabi-Yau threefold has an $SU(3)$ holonomy. } and the contributions from the bosonic degrees of freedom in terms of  $P^3:H^2(X,\IZ)\rightarrow \IZ$ and  $c_2(X)\in H^4(X,\IZ)$ the second chern class of the threefold $X$, are 
the $m_p=\frac{1}{3}P^3+\frac{1}{6}c_2(X)\cdot P-2$ real  moduli of $P $ and the $b_2^-=\frac{2}{3}P^3+\frac{5}{6}c_2(X)\cdot P-1$ dimensional space of anti-self dual  two-forms on $P$. Therefore,
\begin{equation}
c_L=m_p+b_2^-+3 \ ,
\end{equation}
where the last contribution represents the $3$ translation zero modes. 
Similarly, the bosonic contribution to $c_R$ is given by the moduli $m_p$ and the 
$b_2^+=\frac{1}{3}P^3+\frac{1}{6}c_2(X)\cdot P-1$ dimensional space of self-dual  two-forms on $P$, together with the 3 translational zero modes. In addition, $c_R$ also has $f_p=\frac{1}{3}P^3+\frac{1}{6}c_2(X)\cdot P$ fermion contributions as required by supersymmetry which come from $(0,2)$ forms on $P$.
Hence, 
\begin{equation}
c_R=m_p+b_2^+ +3+f_p
\end{equation}
The $c_L,c_R$ central charges can be expressed in terms of $P$ as
\begin{equation}\label{eq:central2}
\boxed{c_L=P^3+c_2\cdot P \ \text{ and } c_R= P^3+\frac{1}{2}c_2(X) \cdot P} \ ,
\end{equation}

In particular, one can note that the right-moving central charge $c_R$ is mapped to the geometric genus $p_g=h^{2,0}(P)$ of the divisor $P$. 
This can by seen by considering the  holomorphic Euler characteristic $\chi(\mathcal{O}_P)$ which by Theorem~\ref{Hrrt} is given by 
\begin{eqnarray}\label{eq:ga}
\chi(\mathcal{O}_P) = \frac{1}{6}P^3 + \frac{1}{12}P\cdot c_2(X)  =p_g+1 \quad \Longrightarrow \quad \boxed{c_R=6(p_g+1)} \ ,  \nonumber \\
\end{eqnarray}
Note that we used the fact that $\chi(\mathcal{O}_P)=p_a+1=p_g-q+1 $ where $p_a$ is the arithmetic genus of $P$ and that the irregularity $q=h^{1,0}(P)=0$, as we saw previously by (\ref{Lht}). 
The left-moving central charge $c_L$ including the center-of-mass contribution can be understood as the topological  Euler characteristic $\chi(P)$:
\begin{equation}\label{eq:chiX}
\chi(P) = c_L = P^3 + P\cdot c_2(X) \ .
\end{equation}

In the case of semi-ample divisors with $P^3=0$ inside a Calabi-Yau threefold, technically the irregularity $q=h^{1,0}(P)$ might not be zero. 

The cases of surfaces with a non-zero irregularity either lead to SUSY enhancement of the 2d worldsheet CFT or describe 6d supergravity strings.

\begin{claim}\label{claimq}
	Any surface with $q>0$ corresponds either to a string of a 6d supergravity compactified on a circle or susy enhancement of the worldsheet CFT with the exception of $\kappa=1$ surfaces without a section \footnote{As described previously this class of surfaces does not describe 6d strings but they can be irregular with  $q\neq 0$. In particular, consider the bundle $L$ isomorphic to the Hodge bundle (whose fiber over $p \in C$ is the 1-dimensional vector space of holomorphic 1-forms on the elliptic fiber $E_p$ over $p$).  The Hodge bundle is trivial if and only if you can find (globally over $C$) a family of nonvanishing holomorphic 1-forms on the fibers $E_p$ which varies holomorphically in $p$. In the case that it is trivial $q=g+1$, otherwise $q=g$ \cite{MR1600388}.  }.
	In particular, 
smooth	K\"ahler irregular surfaces fall in the following classes:
	\begin{itemize}
		\item $\kappa=0$ hyperelliptic surfaces: which have $q=1$, $c_2\cdot P=0$ with $c_R=c_L=6$. This surface gives rise to $\mathcal{N}=(4,4)$ supersymmetry enhancement.
		\item $\kappa=0$ abelian surfaces: which have $q=2$, $c_2\cdot P=0$ with $c_R=c_L=12$.
		This surface gives rise to $\mathcal{N}=(8,8)$ supersymmetry enhancement.
		\item $\kappa=1$ elliptic surfaces (with a section)  over genus $g>0$ curves: which have $q=g$ \footnote{When the surface is a trivial fibration of the form $C\times E$ for $C$ a genus $g$ curve and $E$ an  elliptic curve, $q=g+1$ and hence $c_R=6(g+1), \, c_L=6(g+1)$ since $c_2\cdot (C \times E)=0$.}, $c_2\cdot P\geq0$ with $c_R=6g+\frac{1}{2}c_2\cdot P, \, c_L=6g+c_2\cdot P$.
		These surfaces give rise  to 6d supergravity strings \cite{Kim:2019vuc} on circle.
	\end{itemize}
\end{claim}

The central charge of the first two surfaces can be computed to be  $c_R=c_L=6q$ by similar methods as we did in Section \ref{MSWcentral} but with $h^{1,0}(P)\neq 0$. They both have the same number of left-moving bosons and fermions, $N^B_L=N^F_L=4q$ where the left-moving fermions are induced by the non-zero $h^{1,0}(P)=q$ \cite{Maldacena:1997de}. This is compatible with (4,4) and (8,8) supersymmetry enhancements of the worldsheet CFT.
However, the last case of surfaces gives rise to 6d supergravity strings on a circle. This is because by \cite{Wilson1994,OguisoK} we know that the existence of $\kappa=1$ elliptic surfaces with $c_2\cdot P\ge0$ inside a Calabi-Yau threefold means that the threefold always has an elliptic fiber structure. Since the threefold is elliptic we can invoke the M-theory/F-theory duality and realize the string coming from M5-brane wrapping the elliptic surface over genus g curve $C$ as a D3-brane wrapping  the curve  $C$. As was discussed in \cite{Haghighat:2015ega,Couzens_2017} the central charges for these surfaces can be computed as in Section \ref{MSWcentral} with non-vanishing $h^{1,0}$ with the addition of an emergent $SU(2)_R$ flavor symmetry in the IR. Therefore, the central charges will be the same as in the formula (\ref{eq:central2}) as expected.

An important note is that the geometric genus $p_g$ of the surface $P$ is an integer number which is in accordance with the quantization condition of $c_R \in 6\mathbb{Z}$ as seen in (\ref{eq:ga}) .

The formulas for the $c_R,c_L$ central charges match the one found from field theories in (\ref{eq:central}), which are given by a combination of the cubic and linear Chern-Simons terms evaluated on the charge $q^I$ string. These Chern-Simons terms are the Chern classes of the divisor $P$ which can be express in terms of a bases $\omega_I$ of $H^2(X,\IZ) $ and charges $q^I$ as $P=\sum q^I \omega_I$:
\begin{equation}\label{eq:compare}
P^3 \equiv \int_P c_1^2(P) = C_{IJK}q^Iq^Jq^K \ , \quad P\cdot c_2(X) \equiv \int_P c_2(X)  = C_Iq^I \ .
\end{equation}

Until now we only considered smooth divisors, but there is no reason to assume that the divisor is smooth and in fact there is no simple algebraic criterion to determine smoothness. However, even though the geometric procedure of computing the central charges of Section \ref{MSWcentral} is no longer well-defined, we can still compute the central charge from physics as we did in Section \ref{sec:2} through anomaly inflows. This procedure shows that the central charges are still given by (\ref{eq:central})  and hence by the equation (\ref{eq:compare}), so they can still be expressed geometrically as (\ref{eq:central2}). Unfortunately, there is no  classification for singular K\"ahler surfaces as we saw for smooth ones in Table 	\ref{tb:minimal-S}. However, if we have  a singular nef surface $P$   then it is also semi-ample as we saw previously and hence some multiple  $mP$ for $m \geq 0 $ is smooth. Therefore,  $mP$ is smooth and semi-ample and hence is one of the smooth surfaces described in Table 	\ref{tb:minimal-S}. For example, consider  $mP$ to  be a   smooth K3 surface for  $m>1$. In that case we have that  $(mP)^3=0$, hence $P^3=0$. From eq. (\ref{eq:ga}  ) we can see that $2P^3+c_2\cdot P$ is a multiple of 12, we conclude that $c_2\cdot P \in 12 \mathbb{Z}$ since $P^3=0$.  We also know that  $c_2\cdot mP =24$ for K3 , and therefore we conclude that $m=2$.  Similarly, we note that  $mP$ cannot be a smooth Enriques because in that case $c_2\cdot mP =12$, hence $m=1$.  But we assumed that P is not smooth. 

\subsection{Conditions on 4-cycles and strings}

In this section, we will analyze the geometric condition that arise by considering  semi-ample, ample or very ample divisors in a compact threefold together with their implications as Swampland conditions for the associated  supergravity theory. In particular,  we will relate the conditions on the divisors of supergravity strings with the constraints on the gauge group and the matter content in the bulk supergravity theory. Interestingly, some of the geometric conditions can be interpreted as unitarity constraints on 2d worldsheet CFTs of supergravity strings.

\subsubsection{Geometric conditions on the supergravity strings}
 The various properties that the divisors will need to satisfy will lead us to various consistency conditions for the supergravity when geometrically engineered. Some of the conditions which we will discuss below can also be derived from a field theory analysis of the 5d supegravity theory, and some are new ingredients that do not have  obvious origin in the physics.
Supergravity strings amount to semi-ample (or, equivalently, nef) divisors in a Calabi-Yau threefold. 

As we have discussed above, a smooth semi-ample divisor is a minimal surface with Kodaira dimension $\kappa \ge 0$ (See Table \ref{tb:minimal-S}).
Some bounds on the invariants of these minimal surfaces are listed in Table~\ref{tb:conditions}.
These bounds are discussed in some detail in appendix \ref{appendix}.
 We conjecture that these bounds hold for singular semi-ample divisors as well.

\begin{table}[h!]
	\centering
	\begin{tabular}{|l|}
		\hline
		\\
		1. $P^3\geq 0$, in particular 
		$\left\{
		\begin{matrix}
		P^3=0 &\text{ for } \kappa=0,1\\
		P^3>0 &  \text{ for } \kappa=2\\
		\end{matrix}
		\right.$	\\ 
		\\
		2. $  P\cdot c_2(X)\geq 0$, \  (Theorem\ \ref{thm:miyaoka})
                         \ (with strict inequality if ample)\\ 
                       \ \ \   and $P \cdot c_2(X)\in 2\mathbb{Z}$ (Theorem \ref{Hrrt}).
		\\

		3.	$4P^3-P\cdot c_2(X)+C\ge0 \text{ with } C=\left\{
\begin{array}{cl}
36 &P^3\ \text{even}\\
30 & P^3\ \text{odd}
\end{array}\right.$ \\
  \hspace{.3cm} \text{for smooth surfaces with $\kappa=0,2$} (Theorem~\ref{Nt}) \\ \\
		4. $h^{1,1}(X) \le P^3 + P\cdot c_2(X) -2  $ for $P$ 
smooth and ample 
(Theorem~\ref{I5})
\\ \\
		5. $N_{-2}\le \frac{1}{6}(4P^3 + 5P\cdot c_2(X)) -1  $ for $P$ 
                          big and nef
                         (Theorem~\ref{I6}) 
		\\ \hspace{.5cm}Here, $N_{-2}$ denotes the number of rational $(-2)$ curves on $P$.
\\ \\ \hline
	\end{tabular}
	\caption{Conditions that the surface $P$ needs to satisfy depending on its general properties.}\label{tb:conditions}
\end{table}

In the first inequality, $P^3\ge 0$ follows from Theorem~\ref{Nef} while the subcases follow from Proposition~\ref{prop:kodairaprop}.
The first two inequalities imply that the supergravity strings arising from M-theory on a threefold $X$ have non-negative cubic and linear Chern-Simons terms found in eq.(\ref{eq:compare}).
The fourth bound can be rewritten as $h^{1,1}(X)\leq c_L-2$ and it will precisely match the constraint on the rank of the gauge group of the low-energy 5d theory coming from a constraint on the unitary 2d CFTs living on supergravity strings.
The fifth bound, as we will see in the next section, will turn out to be very useful in constraining the rank of the non-Abelian gauge groups in the low-energy theory engineered by geometry. This inequality does not seem to have an obvious origin from physics.

Another important property is given by the \textit{Hodge index theorem} (Theorem~\ref{HI}). This theorem tells us that  on any smooth divisor $P$, the intersection product  on $H^2(P)$ has signature  $(1,h^{1,1}(P)-1)$. 
This mathematical theorem is interpreted in physics as the condition (\ref{eq:signature}) on the signature of levels of the current algebras in the worldsheet CFT on the supergravity string.

Lastly, there exists an interesting inequality on the Hodge numbers of a Calabi-Yau threefold $X$:

\begin{framed}
\noindent
	\begin{eqnarray}\label{eq:h21}
		-36P^3-80 \le \frac{c_3(X)}{2}=h^{1,1}(X)-h^{2,1}(X)\le 6P^3 + 40 \quad \text{for $P$ very ample}  \qquad
	\end{eqnarray}
\end{framed}
		Moreover, the inequality can be sharpened by replacing the left hand side by $ -80$, $-180$ and  the right hand side by 28, 54 when $P^3= 1,3$ respectively.
The proof of this inequality can be found in \cite{kanazawa2012trilinear}.

In particular, this inequality  does not seem to have an obvious origin from physics but it provides a strong bound on the dimension of the Higgs branch, therefore on the representations of matter hypermultiplets in the 5d effective theory. 

Suppose for example that a threefold $X$ leads to an effective theory with gauge group $G$, and this theory can Higgs to another threefold $X'$ with $h^{1,1}(X')=1$. The gauge group $G$ of the original theory will be broken to $U(1)$ under this Higgsing. The charged hypermultiplets in the original theory parametrize the Higgs branch of the moduli space \cite{Strominger:1995cz} which is a subspace of the complex moduli parametrized by cohomology classes in $h^{2,1}(X')$. This implies the relation 
\begin{equation}
\sum_i{\rm dim}(R_i) -{\rm dim}(G) + 1\le h^{2,1}(X') \ ,
\end{equation}
where $i$ runs over all hypermultiplets and ${\rm dim}(R_i)$ is the dimension of the representation $R_i$ of the $i$-th hypermultiplet.
The resulting 3-fold $X'$ has a single K\"ahler class represented by an ample divisor $P$. According to (\ref{Opt}), the divisor classes $nP$ are very ample when $n\ge 10 $. Using the inequality (\ref{eq:h21}) for $h^{2,1}(X')$, we find a bound on the representations of charged hypermultiplets in the original theory of the 3-fold $X$:
\begin{equation}
\sum_i{\rm dim}(R_i) -{\rm dim}(G) \le36n^3P^3+80 , \text{ where $P$ is ample and } n \geq 10\ .
\end{equation}

\section{Constraints on supergravity theories}\label{sec:constraints}
In this section, we will constrain 5d supergravity theories by using the geometric conditions on semi-ample divisors and the unitary conditions on worldsheet CFTs of supergravity strings presented in the previous sections.

\subsection{$U(1)\times G$ theories}\label{sec:u1g}
The first example is the supergravity theory with $U(1)\times G$ gauge group where $G$ is a product of non-Abelian groups $G=\prod_iG_i$. Without loss of generality, we can choose a basis for the $U(1)$ divisor $H$ such that both its triple intersection and K\"ahler parameter $\phi^0$ are positive. In addition the gauge couplings $h_i$ for non-Abelian symmetries are required to be positive. The effective theory in this basis has
\begin{eqnarray}
	H^3=C_{000}>0 \,, \ \phi^0>0 \quad {\rm and} \quad h_i=h_{i,0}\phi^0>0 \ \ {\rm for \ all} \ G_i \ . 
\end{eqnarray}

The perturbative hypermultiplets carrying the $U(1)$ charge have masses proportional to $\phi^0$ and integrating them out leads to shifts in the Chern-Simons levels. We assume that all such $U(1)$ hypermultiplets are already integrated out. Then the remaining perturbative states are charged only under the non-Abelian group.

There can also be non-perturbative states carrying the $U(1)$ charge: for example, the instanton particles of the non-Abelian gauge group $G$. When the Coulomb branch parameters for $G$ are small enough compared to $h_i$, the BPS instanton state has mass proportional to the gauge coupling, $|m_{\rm inst}|=h_i$ up to a constant factor. This implies that all the BPS particles carrying non-zero $U(1)$ charges have positive electric charge under the $U(1)$ gauge symmetry. The divisor $H$ is the dual to this $U(1)$ gauge symmetry. Therefore the magnetic monopole string with positive charge $q$ on this divisor $H$ is a supergravity string and the divisor $H$ is thus semi-ample. This should be true even if the supergravity theory is not geometrically realized.
Since the string on $H$ is a supergravity string, we can analyze consistency of this string by using the conditions presented in the previous sections and can examine if the bulk gravity theory with the string is consistent or not.

The worldsheet theory on the monopole string of the divisor $H$ should be a $\mathcal{N}=(0,4)$ CFT. Since the $H$ positively intersects the gauge divisors, the worldsheet theory should contain unitary representations of current algebras for $G$. We find that the CFT on a single string with unit magnetic charge $q=1$ contains the current algebras for the bulk gauge group $G_i$ at level $k_i=h_{i,0}$. Then the unitary condition in (\ref{eq:level-c}) puts a bound on the total rank of the non-Abelian gauge group as
\begin{equation}
	\sum_i c_{G_i}=\sum_i \frac{k_i\cdot {\rm dim}G_i}{k_i+h^\vee_i} \le \hat{c}_L \quad \rightarrow \quad \sum_i r_i \le C_{000} + C_0 -3 \ ,
\end{equation}
where $r_i={\rm rank}(G_i)$.
Here we have used the fact that $c_{G_i}$ takes the minimum value $r_i$ when $k_i=1$. In particular, this shows that the rank of the non-Abelian gauge group $G$ in the bulk 5d supergravity theory is bounded from above by the Chern-Simons coefficients $C_{000}$ and $C_0$. 
We remark that this bound generically holds for any 5d supergravity theory with gauge group $U(1)\times G$ regardless of whether it admits geometric construction or not.

For example, suppose that a supergravity theory with $U(1)\times G$ gauge group is Higgsed to a quintic threefold. The Higgsing does not change the Chern-Simons coefficients $C_{000}$ and $C_0$. This implies that the Chern-Simons levels of the original theory before being Higgsed are fixed to be those of the quintic hypersurface $H$, i.e. $C_{000}= H^3=5$ and $C_0=H\cdot c_2=50$. From this, we find a strict bound on the rank of the non-Abelian gauge group $G$
\begin{equation}
	\sum_i r_i \le 52 \ ,
\end{equation}
in any supergravity theory with a single $U(1)$ symmetry which Higgses to a quintic threefold.

When the supergravity theory before being Higgsed has a geometric construction, we can find a stronger bound by using the geometric bound in (\ref{I6}). Note that the non-Abelian symmetry $G$ can remain unbroken, when the K\"ahler parameter $\phi^0$ is turned on, only if the $U(1)$ divisor $H$ and the gauge divisors $E_i$ are glued along rational $(-2)$ curves in $H$ and the fibers in $E_i$. Also $H$ must be glued to all $E_i$ divisors in order that the low-energy theory has gauge couplings $h_i$ with proper signs. This imposes a bound $\sum_i r_i \le N_{-2}$ on the rank of the non-Abelian group with respect to the number of $(-2)$ curves in $H$. Therefore, the bound (\ref{I6}) on $N_{-2}$ tells us that
\begin{equation}\label{eq:rank-bound}
	\sum_i r_i \le \frac{2}{3}H^3 + \frac{5}{6}H\cdot c_2(X)-1 = 44 \ .
\end{equation}
for supergravity theories admitting M-theory construction on Calabi-Yau 3-fold that reduce to a quintic threefold after Higgsing.

In Appendix \ref{app:examples}, we present a number of concrete constructions of compact Calabi-Yau threefolds that Higgs to a quintic threefold. These geometries engineer the supergravity theories with gauge group $U(1)\times SU(2)$ or $U(1)\times SU(3)$. It is obvious that the above bound (\ref{eq:rank-bound}) is consistent with these examples. One can check that divisors contained in these geometries and the corresponding monopole strings satisfy all the conditions we listed in the previous sections.

The first example is for the supergravity theory with $U(1)\times SU(2)$ gauge group coupled to $N_f=9$ $SU(2)$ fundamental hypermultiplets. The threefold consists of two divisor classes $H$, the proper transform of the hyperplane class of the quintic, and $E$, the exceptional divisor of the blowup. The triple intersections of $H$ and $E$ classes are given in (\ref{eq:triple-intersections-SU2-9}). Two divisors are glued along a $(-2)$ curve in $H$ and the fiber class $r$ in $E$. In this example, the K\"aher cone is generated by $H$ itself and $H-E$.
 The supergravity strings are then the M5-branes wrapping any 4-cycles $D$ which can be written as
\begin{equation}
	D = mH + n(H-E) \ , \quad m,n\ge0 \ .
\end{equation}
The central charges of the supergravity strings are
\begin{eqnarray}
	\hat{c}_L &=& m(5m^2+15mn+9n^2)+50m+36n-3 \ , \nonumber \\
	\hat{c}_R &=& m(5m^2+15mn+9n^2)+25m+18n-6 \ .
\end{eqnarray}
One can easily see that the central charges are positive and $\hat{c}_R \in 6\mathbb{Z}$ for the non-trivial supergravity strings as expected.
On generic points of the K\"aher moduli space, the gauge symmetry is broken to $U(1)\times U(1)$.
We checked that the signature of the levels of $U(1)\times U(1)$ current algebras, which include the center-of-mass sector, in the supergravity strings is always ${\rm sig}(k_{IJ})=(1,1)$. This is consistent with the condition (\ref{eq:signature}) on the signature. This result is guaranteed by the Hodge index theorem \ref{HI} for the semi-ample divisors in this geometry. All other geometric conditions are surely satisfied. 

When we turn on only the K\"ahler parameter of the $H$ class, then the $SU(2)$ gauge symmetry remains unbroken. In this case, the 2d CFT on the monopole string of the $H$ class carries a $U(1)$ current algebra at level $5$ in the right-moving sector and a $SU(2)$ current algebra at level $1$ in the left-moving sector. The central charge of the $SU(2)$ current algebra in the worldsheet is $c_{SU(2)}=1$. So the inequality $c_{SU(2)} \le \hat{c}_L$ in eqn. (\ref{eq:level-c}) is satisfied.

More examples of supergravity theories with $U(1)\times SU(2)$ gauge symmetry  are given in Appendix \ref{app:SU2}.

The second example is the supergravity theory of $U(1)\times SU(3)$ gauge symmetry with $N_f=11$ fundamental hypermultiplets. The CY$_3$ geometries for this theory are constructed in Appendix \ref{app:SU3}. Each threefold is labelled by an integer $0\le n\le 3$ and corresponds to the $SU(3)$ gauge group at Chern-Simons level $\kappa=-\frac{3}{2}+n$. The threefold consists of three surfaces, $H, E_1$ and $E_2$. The divisor $H$ is the proper transform of the hyperplane class of the quintic and $E_1$ is a Hirzebruch surface $\mathbb{F}_{5+n}$ with 11 blowups\footnote{Our construction exhibits this surface as an 11-fold blowup of $\IF_0$ or $\IF_1$ depending on the parity of $n$, but by blowing down the half-fibers disjoint from the section where the two components are glued, we get $\mathbb{F}_{5+n}$.} and $E_2$ is a Hirzebruch surface. Two $(-2)$ curves in $H$ are each glued to a fiber in $E_1$ and another fiber in $E_2$. Two surfaces
$E_1$ and $E_2$ are glued along the section $e^2=-(n+5)$ in $E_1$ and a rational curve $C^2=n+3$ in $E_2$.
The triple intersections of three surfaces are given in eqn. (\ref{mostint}).

Let us consider the case with $n=0$. The K\"ahler cone is generated by $H, H-E_1-E_2$ and $2H-2E_1-E_2$. The supergravity strings come from the M5-branes wrapping linear combinations of these generators with non-negative integer coefficients. The central charges $(\hat{c}_L,\hat{c}_R)$ for these three generators are $(52,24), \, (33,12)$ and $(79,42)$ respectively. As expected since the (self-)triple intersections of the generators are non-negative, all supergravity strings have positive central charges. We also checked that the signature of the level of the current algebras is ${\rm sig}(k_{IJ})= (1,2)$ for all three generators which is in accordance with the condition (\ref{eq:signature}).

The low-energy theory has the $SU(3)$ gauge symmetry enhancement along the K\"ahler moduli space of the K\"ahler parameter for $H$. The M5-brane wrapping the divisor $H$ gives rise to a monopole string hosting in the left-moving sector a level $1$ current algebra for the $SU(3)$ symmetry. The unitary condition $c_{SU(3)}\le \hat{c}_L$ for this monopole string is therefore satisfied with $c_{SU(3)}=2$ and $\hat{c}_L=52$.

\subsection{Abelian gauge theories}
Now consider a generic point on the Coulomb branch of the moduli space in a supergravity theory engineered in M-theory on a CY$_3$. The gauge symmetry $G$ in the 5d supergravity is fully broken to its Cartan subgroup $U(1)^{{\bf r}+1}$. There are a set of basis 4-cycles $P_I$ with $I=0,1,\cdots, {\bf r}$ for the $U(1)^{{\bf r}+1}$ gauge group. We claim that the holomorphic surfaces $P_I$ can always be chosen to be semi-ample divisors in the 3-fold. In other words, all Abelian gauge groups in the low energy effective theory can be represented by a set of ${\bf r}+1$ semi-ample divisors that are part of the K\"ahler cone generators. Since they are semi-ample, the corresponding strings are all supergravity strings.

The effective Abelian theory is characterized by the triple intersections $C_{IJK} = \int_X P_I \cdot P_J \cdot P_K$ and the second Chern classes $C_I = \int_X P_I \cdot c_2(X)$. From the fact that all $P_I$ are {\it nef} and semi-ample divisors, one finds that the triple intersections and the second Chern classes are all non-negative. We propose 
\begin{equation}
	C_{IJK} \ge 0 \ , \quad C_I \ge 0 \quad {\rm for \ all} \ I, J, K \, ,
\end{equation}
from the properties of semi-ample divisors.
The inequalities for $C_{III}$ and $C_I$ are obvious by the definition of semi-ample divisors. Also, $C_{IIJ}$ with $I\neq J$ is the sum of intersection numbers $P_I\cdot C$ between a semi-ample divisor $P_I$ and curves $C$ at the intersection $P_I \cap P_J$, which tells us that $C_{IIJ}\ge0$. Similary, $C_{IJK}$ is the sum of intersection numbers between the divisor $P_I$ and curves at the intersection $ P_J \cap P_K$, and it needs to be non-negative for the semi-ample divisors.

In addition, the surfaces $P_I$ must satisfy the Hodge index theorem. For ${\bf r}=1$ cases, for instance, the Hodge index theorem or the relation (\ref{eq:signature}) says that the signature of the levels $k_{IJ}$ with $I,J=0,1$ should be $(1,1)$ for the worldsheet CFT on wrapped M5-brane over each $P_I$. One can then deduce the following two conditions on the triple intersections from the M5-branes wrapping once on $P_0$ and $P_1$ respectively :
\begin{equation}
	C_{000}C_{001} \le C_{011}^2 \ , \quad C_{111}C_{011} \le C_{001}^2 \ .
\end{equation}

\subsection{Generic gauge theories}

We will now turn to supergravity theories coupled to generic gauge groups.
If these theories can be geometrically engineered, then we can constrain them by using conditions on divisors in the 3-fold as follows.

Let us consider a 3-fold $X$ and the low-energy theory at a special submanifold on the K\"ahler moduli space of $X$ where some Abelian symmetries enhance to non-Abelian symmetries $G=\prod_iG_i$. More precisely, we are interested in the effective theory in the moduli space where all the K\"ahler parameters $\phi_\alpha$ for $U(1)$ symmetries are taken to be large, while the K\"ahler parameters $\phi_i$ for some non-Abelian symmetries $G$ are turned off. If the non-Abelian symmetry $G$ in the low-energy theory remains unbroken even after integrating out all matters charged under the Abelian symmetries, then we say that M-theory compactified on $X$ at low-energy is described by the supergravity theory with gauge group $G$ times multiple Abelian factors. We shall now assume this and constrain such effective theories.

We first conjecture that all the 4-cycles $P_\alpha$ for Abelian gauge groups in $X$ can be chosen to be semi-ample divisors. So there exists a basis where all $P_\alpha$'s for $U(1)$ gauge groups are semi-ample. In this basis we find the following conditions on the Chern-Simons levels,
\begin{equation}\label{eq:generic-C}
	C_{\alpha\beta\gamma}\ge0\ , \quad C_\alpha \ge 0 \quad {\rm for \ all} \ \alpha,\beta,\gamma \ ,
\end{equation}
where $\alpha,\beta,\gamma$ denote the indices for the Abelian gauge groups.
These conditions again follow from the fact that divisors $P_\alpha$ are semi-ample and they non-negatively intersect all effective 2- and 4-cycles in $X$. 

The requirement for the non-Abelian symmetry $G_i$ unbroken imposes non-trivial constraints on the intersection structure between the Abelian divisors $P_\alpha$ and non-Abelian gauge divisors $E_i$.
First, since we want to preserve the non-Abelian symmetry $G$ on the moduli space of the K\"ahler parameter $\phi^\alpha$ for $P_\alpha$, the triple intersections $C_{\alpha\beta i}$ necessarily vanish. Otherwise the corresponding Chern-Simons interaction (partially) breaks the symmetry $G$. This condition $C_{\alpha\beta i}=0$ should be true even after we turn on small K\"ahler parameters $\phi^i$ for the non-Abelian gauge divisors $E_i$ because the massive states sitting in some representations of the non-Abelian symmetry that is weakly broken by $\phi^i$ after integrated out cannot induce Chern-Simons terms with coefficient $C_{\alpha\beta i}$.

Also the gauge couplings of the non-Abelian groups need to be positive. This forces 
\begin{equation}\label{eq:generic-h}
	h_{i,\alpha} \ge 0 \ ,
\end{equation}
for all $i$ and $\alpha$.
From this, one can deduce more conditions on the Chern-Simons levels when $X$ is fully resolved. Let us turn on small K\"ahler parameters $\phi^i$ for non-Abelian gauge divisors $E_i$ and assume $\phi^i\ll \phi^\alpha$. The positivity of the gauge couplings (\ref{eq:generic-h}) is then translated into 
\begin{equation}
	C_{\alpha ii} \le 0 \quad {\rm for \ all} \ i \ , \quad C_{\alpha ij} \ge 0 \quad {\rm for\ all } \ i\neq j \ .
\end{equation}

\section{Conclusion}\label{sec:conclusion}
In this paper, we have proposed conditions that allowed effective theories of quantum gravity in five-dimensions must satisfy. The essential idea was to relate unitarity of magnetic monopole strings that only appear in a gravitational theory to the conditions on the topological data in the supergravity effective action. When the low-energy theory is constructed in M-theory on compact Calabi-Yau 3-fold, such strings are associated to special 4-cycles called semi-ample (or nef) divisors in the 3-fold. We have shown that algebraic conditions on these divisors put bounds on the number of massless degrees of freedom in the gravity theory and constrain their interactions in the effective action. For generic supergravity theories which may not admit geometric construction, we obtained a weaker bound on the rank of the gauge symmetry in the gravity theory.

We have seen that admissible 5d supergravity theories in M-theory compactification are constrained by a series of necessary conditions for compact Calabi-Yau 3-folds and also those for algebraic surfaces theirof. Some of these conditions, for example the Hodge index theorem for algebraic surfaces, were shown to be related to the unitarity of strings in the supergravity theories. Other geometric conditions presented in this paper could also have physical interpretations, and studying them will possibly provide us a deeper understanding for more general properties of quantum gravity in 5d. We leave this for future work. 

Semi-ample divisors in 3-folds and associated monopole strings have turned out to play a central role in examining consistency of gravitational theories. Though semi-ampleness is a well-defined concept in the intersection theory in mathematics, it is a practically non-trivial task to correctly identify which divisor is (semi-)ample. It is because of this that we need to know all the intersections of a divisor with every 2-cycle in the Mori cone, or equivalently we need to know charges of all BPS particles in the gravity theory. There is currently no systematic approach for achieving this in general 3-folds and supergravity theories. Nonetheless, we can systematically extract topological data of 2d CFTs on monopole strings such as anomalies. Perhaps semi-ampleness is fully encoded in such 2d CFT data so that we can precisely isolate supergravity strings without relying on a geometric construction, but this would demand more careful studies on algebraic surfaces in geometry and monopole strings in 5d supergravity theories.

\section{Acknowledgments}
We would like to thank Thomas W. Grimm, Seok Kim, Thomas Peternell, and David Tong for valuable discussions. HK would like to thank Harvard University for hospitality during part of this work.
The research of HK is supported by the POSCO Science Fellowship of POSCO TJ Park Foundation and the National Research Foundation of Korea (NRF) Grant 2018R1D1A1B07042934. The research of HCT and CV is supported in part by the NSF grant PHY-1719924 and by a grant from the Simons Foundation (602883, CV).  The research of SK is supported by the NSF grant DMS-1802242.

\appendix

\section{Appendix A: Mathematical facts and proofs}\label{appendix}

In this section, we describe the concepts and results which we need about divisors and linear systems on surfaces and threefolds.   
Our motivation is to understand the extent to which numerical conditions on divisor classes on compact Calabi-Yau threefolds do or do not guarantee the existence of a smooth surface.  While no numerical criterion exists, there is a rich classical theory in algebraic geometry providing many results in that direction which will be useful for us. 

A good general reference which touches on many of these issues is the book \cite{Lazarsfeld:2004pag}.

\subsection{Definitions}\label{sec:defs}
We begin by stating the relevant definitions and notions from algebraic geometry.  We consider divisors $P$ on smooth projective varieties $X$ of dimension $n$.   A divisor can be expressed in terms of its irreducible components $P_i$ as $P=\sum_i n_iP_i$ with $n_i\in \IZ$.   Our primary interest is $n=2$ or 3.  

\begin{defn}\label{effective2}
The divisor $P$ is \emph{effective} if all $n_i\ge 0$.  
\end{defn}

\begin{defn}\label{effective}
The divisor of a nonzero meromorphic function $f$ on $X$ is given by $(f)=(f)_0-(f)_\infty$, where $(f)_0$ is the divisor of zeros of $f$ including multiplicity and $(f)_\infty$ is the divisor of poles of $f$ including multiplicity.  
\end{defn}

To a divisor $P$, we associate the sheaf $\CO_X(P)$ of meromorphic functions $f$ on $X$ with $(f)+P$ effective.  By convention, the 0 function is also a section of $\CO_X(P)$.  Such an $f$ can be viewed as a \emph{holomorphic} section $s_f$\footnote{This is non-standard notation, introduced to expedite the discussion.} of $\CO_X(P)$.  For a general $f$ (i.e.\ one not necessarily satisfying $(f)+P$ effective), the corresponding $s_f$ might only be a meromorphic section of $\CO_X(P)$.  We can  equivalently think of a meromorphic function $f$ on $X$ as either a function or a meromorphic section $s_f$ of $\CO_X(P)$.  Conversely, identifying a nonzero meromorphic section $s$  of $\CO_X(P)$ with a meromorphic function $f$ on $X$ (so that $s=s_f$), we define the divisor $(s)$ of $s$ as $(f)+P$.  In particular $s$ is a holomorphic section of $\CO_X(P)$ if and only if $(s)$ is effective. 
This observation leads to conclude that we can always  find an effective divisor associated to $P$ if and only if $h^0(X,O_X(P))\neq 0 $ given by Proposition \ref{eff}. 

\begin{defn}\label{linsys}
Two divisors $P$ and $P'$ are \emph{linearly equivalent}, denoted $P\sim P'$, if there exists a nonzero meromorphic function $f$ on $X$ with $P'=(f)+P$.  The \emph{complete linear system} $|P|$ of $P$ is the set of all effective divisors linearly equivalent to $P$.
\end{defn}

\begin{defn}\label{nef}
A divisor $P$ is \emph{nef} if $P\cdot C\ge0$ for all curves $C\subset X$. 
\end{defn}
The term nef is in part intended as an acronym for ``numerically eventually free".  To say that a divisor is ``eventually free" means that some positive multiple is base point free:

\begin{defn}\label{semidef}
The linear system $|P|$ is called \emph{base point free} when the intersection of all the divisors in $|P|$ is empty.	
	A divisor $P$ in $X$ is called 	\emph{semi-ample}  if the linear system $|mP|$ is base point free for some $m \in \mathbb{N}$.
\end{defn}

A base point free linear system $|P|$ defines a mapping of $X$ to projective space
\begin{equation}\label{eq:maptop}
\phi_{|P|}:X\to \IP^N,\qquad \phi_{|P|}(x)=(s_0(x),\ldots,s_N(x)).
\end{equation}
In (\ref{eq:maptop}), $\{s_0,\ldots,s_N\}$ is a basis for $H^0(X,\CO_X(P))$.  The map $\phi_P$ depends on the choice of basis, but is well-defined up to a linear change of homogeneous coordinates in $\IP^N$.

If $|P|$ is not base point free, then $\phi_{|P|}$ is not defined precisely at the base points.  But we still get a rational map $\phi_{|P|}:X--\to \IP^N$ whenever $|P|$ is not empty.

\begin{defn}
A divisor $P$ is \emph{very ample} if $|P|$ is base point free and the corresponding map $\phi_{|P|}$ is an embedding. The divisor $P$ is \emph{ample} if $mP$ is very ample for some $m\in \IN$.
\end{defn}
We can always determine if a divisor is ample using the numerical criterion given by Theorem \ref{semi-ample}. Moreover, we can always pass from an ample to a very ample divisor through Theorem \ref{Opt}.

\begin{defn}\label{holdiv}
	The \emph{holomorphic Euler characteristic} of a divisor $P$ in $X$ is the alternating sum 
\begin{eqnarray}
\chi(O_X(P))=\sum^3_{i=1}(-1)^ih^i(O_X(P)),
\end{eqnarray}
where as usual $h^i(O_X(P))=\dim H^i(O_X(P))$.  The holomorphic Euler characteristic of $X$ is defined as $\chi(\CO_X)=1-h^{1,0}(X)+h^{2,0}(X)-\ldots+(-1)^nh^{n,0}(X)$.
\end{defn}

\begin{defn}\label{def:big}
A divisor $P$ in $X$ is \emph{Big} if $h^0(X,\CO_X(mP))\ge c m^n$ for some $c>0$ and all $m\ge m_0$.
\end{defn}

\noindent
\textbf{Remark.} By Riemann-Roch we have $\chi(\CO_X(mP))\sim (P^n/n!)m^n$.    If follows immediate that for divisors $P$ satisfying the vanishing condition $H^i(X,\CO_X(mP))=0$ for $i>0$ and $m\ge m_0$, $P$ is big if and only if $P^n>0$.  Vanishing theorems which imply such vanishing conditions will be discussed in Section~\ref{subsec:proofs}.

\vspace{0.5 cm}
Let $K_X$ be the canonical bundle of $X$ and $P_m(X)=h^0(O_P(mK_X))$ be the m-th plurigenus\footnote{Note that $m=1$ is the geometric genus $h^{2,0}(X)$.} of X.
\begin{defn}
The \emph{Kodaira dimension} $\kappa$ of 
a smooth\footnote{Singular surfaces do not have a canonical bundle in general.  However, effective divisors in a smooth threefold are Gorenstein, hence their dualizing sheaves are again line bundles.  We can generalize the notion of Kodaira dimensions to these surfaces if desired, but there is no known classification.  We do not pursue this point further.} surface 
$X$ 
is defined as follows.
\begin{equation}\label{kodaira}
\kappa(X)=min\{k| \ \ \frac{P_m(X)}{m^k} \text{ is bounded }\}
\end{equation}
When all plurigenera vanish we say $\kappa(X)=-\infty $.

\vspace{0.5 cm}
We can similarly associate a Kodaira dimension to any line bundle $L$ on $X$.

\begin{equation}
\kappa(X,L)=\left\{\begin{array}{ll}
-\infty & H^0(X,L^n)=0\ {\rm for\ all\ }n\ge 1\\
\mathrm{sup}(\{\dim\phi_{L^n}(X)\mid n\ge1\})&{\rm otherwise} 
\end{array}\right.
\end{equation}

We can recover \ref{kodaira} when $L=K_X$ i.e. $\kappa(X):=\kappa(X,K_X)$.

\end{defn}

\smallskip
We now specialize to smooth projective surfaces, which we will denote by $S$ instead of $X$.  We use the standard notation and terminology of classical algebraic geometry.

\begin{defn}
The \emph{geometric genus} $p_g(S)$ of $S$ is the dimension of $H^0(S,K_S)\simeq H^{2,0}(S)$.  The \emph{irregularity} $q(S)$ of $S$ is the dimension of $H^{1,0}(S)$.  The \emph{arithmetic genus} $p_a(S)$ is defined as $p_g(S)-q(S)$.
\end{defn}
In particular, $\chi(\CO_S)=1+p_a(S)$.

When $S$ is clear from context, we simply denote these by $p_g$, $q$, and $p_a$ respectively.  

\begin{defn}
A smooth projective surface $S$ is \emph{regular} if $q=0$.  The surface $S$  is \emph{irregular} if $q>0$.
\end{defn}
Since $b_1(S)=2q$, to say that $S$ is regular is equivalent to the topological condition $b_1(S)=0$.

\begin{defn}
A \emph{$(-1)$-curve} is a curve $C\subset S$ with $C$ isomorphic to $\IP^1$ and $C^2=-1$.  A surface $S$ is \emph{minimal} if it has no $(-1)$-curves.
\end{defn}
By the adjunction formula, equivalently $C$ is a $-1$-curve if and only if $C^2=K_S\cdot C=-1$.

\smallskip
Any $(-1)$-curve $C\subset S$ can be blown down to a smooth surface.  This means that we can find a smooth surface $S_1$ with a point $p\in S_1$ and a holomorphic mapping $f:S\to S_1$ such that $f(C)=p$ and $f$ restricts to an isomorphism of $S-C$ to $S_1-p$.  If $S_1$ is not minimal, then it contains a $-1$ curve, which can be blown down to a surface $S_2$.  It can be shown that this process terminates after finitely many steps and we wind up with a minimal surface $S_n$ and a holomorphic birational map $S\to S_n$ which blows down $n$ $(-1)$-curves in succession.

\smallskip
In this way, the classification of compact K\"ahler surfaces is reduced to the classification of minimal compact projective surfaces. 
In particular, in our case we are only interested in minimal surfaces because a surface with  a $(-1)$ curve can never be nef. 
 We will futher restrict to the classification of compact algebraic surfaces, since all of our supergravity strings arise from wrapping surfaces which are algebraic instead of being merely K\"ahler, as will be explained in Appendix~\ref{subsec:theorems}.  Kodaira's classification of minimal compact algebraic surfaces, organized by Kodaira dimension $\kappa$, is given in Table~\ref{tb:minimalsurf}.
\begin{table}[h!]
	\centering
	\begin{tabular}{|c|c|}
		\hline
		$\kappa$ & minimal projective algebraic surface $S$ \\
		\hline 
		$-\infty$ & $\mathbb{P}^2$, ruled surface   \\ 
		\hline
		$0$ & $K3$, Enriques, hyperelliptic, abelian surface  \\ 
		\hline
		$1$ & minimal elliptic surface  \\ 
		\hline
		$2$ & minimal surface of general type  \\ 
		\hline
	\end{tabular}
	\caption{The first column is  the Kodaira dimension  $\kappa$ of the surface S. The second column presents the Enriques-Kodaira classification of minimal projective algebraic surfaces.}
	\label{tb:minimalsurf}
\end{table}

The first class of surfaces with $\kappa=-\infty$ are ruled surfaces or $\mathbb{P}^2$, which are never nef divisors in a Calabi-Yau threefold. 
In particular, a ruled surface is a $\IP^1$ bundle over a smooth curve $C$, which can have any genus $g$.  If $g=0$, the ruled surfaces are rational and are precisely the Hirzebruch surfaces $\IF_n$.  The ruled surfaces over curves of genus $g>0$ have continuous complex structure moduli and there is no standard notation for them.  We sometimes denote a ruled surface in one of these continuous families by $\IF_n^g$.

The next three cases all represent semi-ample divisors. Enriques surfaces are regular algebraic surfaces which are $\mathbb{Z}_2$ quotients of K3. Hyperelliptic surfaces or bi-elliptic surfaces are finite abelian group quotients of a product of elliptic curves. Abelian surfaces are tori and all other elliptic surfaces other than those with $\kappa=-\infty$ or 0 have $\kappa=1$.

\subsubsection{Ruled and Rational Surfaces with $\kappa=-\infty $}\label{Ap:Ruled}
Ruled and rational surfaces with $\kappa=-\infty $  and their blow ups  are never semi-ample divisors in  a Calabi-Yau threefold. Obviously, if we consider any surface with blowups it will automatically not be semi-ample because it will always contain a rational $(-1)$ curve with $K\cdot C=-1$ coming from the blow up as mentioned above. Therefore, it is enough to consider minimal such surfaces and hence $\mathbb{P}^2$ or $\mathbb{F}_n^g$.

Hirzebruch surfaces $P$ over a curve of  genus $g>1$ have $P^3=8(1-g)<0$ therefore they are not ample. 
As for $g=0$, the cohomology of the ordinary Hirzebruch surface $\mathbb{F}_n$ is generated by the section $e$  with $e^2=-n$ and $f$ a fiber. We can  inspect the intersection of the canonical divisor $K=-2e-(n+2)f$ with any section $h$ satisfying $h^2=n$, which gives us $K\cdot h= -(n+2)<0$. For $g=1$ we have $K=-2e- nf$, and now $K\cdot h=-n<0$ for $n>0$. Finally, if $n=0$ and $g=1$ we have $K=-2e$ and hence $K \cdot f =-2<0$.
Therefore, any ruled surface cannot be semi-ample. Lastly, $\mathbb{P}^2$ has a canonical bundle which satisfies $K=-3\ell$ for the class $\ell^2=1$ with $K \cdot \ell=-3<0$. We conclude that smooth projective surfaces with $\kappa=-\infty $ are never semi-ample.

\subsection{Theorems}\label{subsec:theorems}

In this section, we let $X$ be a smooth projective variety of dimension $n$.   In several situations, we will specialize to the case where $X$ is a Calabi-Yau threefold and add something more precise.  We will always assume that such a Calabi-Yau has $SU(3)$ holonomy, so that $H^{1,0}(X)=H^{2,0}(X)=0$.

\begin{prop}\label{eff}
Given a divisor $P$, there is a 1-1 correspondence between $|P|$ and elements of the projective space $\IP(H^0(X,\CO_X(P))$.  In particular, in which case $|P|$ is a projective space of dimension $h^0(X,\CO_X(P))-1$.
\end{prop}
This is a well-known foundational result (e.g.\ \cite{GH:1994pag}) but we provide a proof to fix ideas.  

\smallskip\noindent{\textbf{Proof}}.
A holomorphic section $s\in H^0(X,\CO_X(P))$ can be written as $s=s_f$ for $(f)+P$ effective.  Thus $(s)=(f)+P\in |P|$.  For any nonzero constant $c$ we have $(cs)=(s)$.  Thus the assignment $s_f\mapsto (f)+P$ induces a map $\IP(H^0(X,\CO_X(P))\to |P|$.

In the other direction, let $P'\in |P|$.  Let $f$ be such that $P'=(f)+P$.  For this $f$ we have $(s_f)=P'$, which was assumed effective.  Thus $s_f$ is holomorphic, i.e.\ gives a section of $H^0(X,\CO_X(P))$.  If instead we choose a different $f'$ with $P'=(f')+P$, then $(f'/f)=(P'-P)-(P'-P)=0$.  It follows that
$f'/f'$ is a holomorphic nonvanishing function on $X$, which must be constant since $X$ is compact.  So $f$, hence $s_f$, is unique up to scalar and we have defined an inverse map $|P| \to \IP(H^0(X,\CO_X(P))$.

\begin{thm}\label{thm:bertini} (\textbf{Bertini's Theorem})
The general member of a base point free linear system is smooth.
\end{thm}

In particular, if $X$ is a Calabi-Yau threefold and $P$ is semi-ample, then we can find a smooth surface in the linear system $|mP|$ for some $m\ge 1$.

\begin{prop}\label{prop:sanef} Semi-ample divisors are nef.\end{prop}

Thus the nef condition is a purely numerical condition on a divisor which is automatically satisfied if it semi-ample, i.e.\ ``eventually free".  

\smallskip\noindent
\textbf{Proof:} Suppose $|mP|$ is base point free.  Let $C$ be an irreducible curve, and $p \in C$.  Since $p$ is not a base point of $ |mP|$, there is a divisor D in $|mP| $ not containing $p$. This implies that $ D$ does not contain $C$, hence $ D\cdot C = (mP)\cdot C \ge 0 $ and finally $P\cdot C \ge 0$.

\smallskip\noindent
Recall the notion of an ample divisor from Appendix~\ref{sec:defs}.

\begin{thm}\label{semi-ample}
(\textbf{\ Nakai–Moishezon criterion for ampleness})
	A divisor $D$ in $X$ is ample  iff  $D^k\ \cdot V > 0 $ for all irreducible subvarieties $V \subset X$, where $k$ is the dimension of $V$.  In particular, if $X$ is a (Calabi-Yau) threefold, $D$ is ample iff the following  three conditions hold: $D^3 > 0$, $D^2\cdot S>0$ for all irreducible surfaces $S\subset X$, and $D\cdot C>0$ for all irreducible curves $C\subset X$.
\end{thm}

The Nakai-Moishezon criterion implies that the ampleness of $D$ only depends on the class  $[D]\in H^2(X,\IR)$.  The cohomology classes of ample divisors span a cone in $H^2(X,\IR)$, the \emph{ample cone}  or \emph{K\"ahler cone} $\mathcal{K}(X)\subset H^2(X,\IR)$.   The reason for the interchangeable terminology is that the line bundles associated to ample divisors are precisely the line bundles which admit K\"ahler metrics.  We will describe the cone spanned by the ample divisor classes as the K\"ahler cone to match usage in physics.

While the Nakai-Moishezon condition is a purely numerical condition, it is not completely satisfactory for our purposes since we have to know \emph{all} surfaces $S\subset X$ in order to implement the criterion.  As we will see presently, it is easier to work with the nef cone, which gives us almost as much information anyway.

\smallskip
The following theorem shows that the nef cone $\mathrm{Nef}(X)$ generated by nef divisors is the closure $\overline{\mathcal{K}(X)}$ of the ample cone, as we just replace ``$>$" in the Nakai-Moishezon criterion with ``$\ge$". 

\begin{thm}\label{Nef} 	(\textbf{Kleiman's Theorem} \cite{Kleiman:1966na})
Let $D$ be a nef divisor on $X$. Then for any  subvariety $V$ of $X$ we have $D^k\cdot V\geq0 $ where $k$ is the dimension of $V$. 
In particular, if $X$ is any threefold, such as a Calabi-Yau threefold, we see that $D^3\geq 0 $. \end{thm}

Said differently, Kleiman's Theorem says that $\mathrm{Nef}(X)=\overline{\mathcal{K}(X)}$ is dual to the Mori cone $M(X)\subset H_2(X,\IR)$, the cone generated by the classes of all irreducible curves $C\subset X$.  The conclusion of Kleiman's theorem holds for semi-ample divisors, since semi-ample divisors are nef by Proposition~\ref{prop:sanef}.  

In general, the determination of the K\"ahler cone $\mathcal{K}(X)\subset\mathrm{Nef}(X)$ is more subtle.  But it can be shown that the ample cone is the interior of the nef cone \cite{Kleiman:1966na}.  In particular, if the nef cone is known to be a polyhedral cone generated by finitely many nef divisors, then this fact determines the ample cone.

\smallskip\noindent
\textbf{Remark.} We have assumed that $X$ is projective throughout this section, so these results do not apply to local Calabi-Yau threefolds.  To see the issue, suppose that $P$ is a smooth surface in a compact Calabi-Yau $X$.  To say that $P$ is nef means that $P\cdot C\ge0$ for all curves $C\subset X$, \emph{not just those contained in $P$}.  In the local case, all compact curves are either contained in $P$ or a deformation of curves contained in $P$, hence homologous to curves in $P$.  So the condition becomes $P\cdot C\ge0$ for all curves $C\subset P$ (as was studied in the context of 5D SCFT \cite{Jefferson:2018irk}), which is a substantially weaker condition than requiring that $P\cdot C\ge0$ for all curves $C\subset X$ for any given compact Calabi-Yau $X$ containing $P$.

\smallskip
For simplicity, we only state the Hirzebruch-Riemann-Roch theorem for Calabi-Yau threefolds.
\begin{thm}
	\textbf{(Hirzebruch-Riemann-Roch theorem) }\label{Hrrt}
The holomorphic Euler characteristic of a divisor $P$ in a Calabi-Yau threefold $X$ is given by
\begin{eqnarray}
\chi(O_X(P))=\frac{1}{6}P^3+\frac{1}{12}P \cdot c_2(X)
\end{eqnarray}
\end{thm}

If $H^k(X,\CO_X(P))=0$ for all $k>0$, then $\dim |P| = \dim H^0(X,\CO_X(P))-1=\chi(\CO_X(P))-1$, and we can compute the dimension of our moduli space of surfaces $|P|$ very simply by Hirzebruch-Riemann-Roch.  We now give a few theorems which guarantee these vanishings of cohomology. 

\begin{thm}
	\textbf{(Kodaira Vanishing theorem \cite{kovcs2000logarithmic})}\label{Kvt}
	Let $P$ be an ample divisor on a smooth projective variety $X.$ Then
	$H^i(X,K_X(P))=0$ for any $i>0$.  In particular, if $X$ is Calabi-Yau we have $H^i(X,\CO_X(P))=0$ for any $i>0$.
\end{thm}
 
Since $\mathrm{Nef}(X)$ is the closure of $\mathrm{Amp}(X)$, one might hope that the desired vanishing holds for nef divisors, but that is not true in general.   A slight strenghtening of the nef hypothesis works which is more general than ample.

\begin{thm}
	\textbf{(Kawamata-Viehweg vanishing\cite{kovcs2000logarithmic}) }\label{Kvt}
	Let $P$ be a nef and big divisor on $X.$ Then
	$H^i(X,K_X(P))=0$ for any $i>0$.  In particular, if $X$ is Calabi-Yau we have $H^i(X,O_X(P))=0$ for any $i>0$.
\end{thm}

Specializing to a Calabi-Yau threefold for definiteness, we see that $\chi(\CO_X(mP))$ grows like $(P^3/6)m^3$.  If $P^3>0$, this is close to the condition for being big, but is not the same since  $\chi(\CO_X(mP))$ is not the same as $\dim H^0(X,\CO_X(mP))$ in general.  To conclude the required growth of $\dim H^0(X,\CO_X(mP))$, the growth of $\dim H^2(X,\CO_X(mP))$ must be controlled for $i>0$.  This can be done:

\begin{prop} \label{prop:big}
Suppose $X$ is a Calabi-Yau threefold, and $P$ is nef and satisfies $P^3>0$.  Then $P$ is big.
\end{prop} 

Since semi-ample divisors are nef, it follows immediately from the Kawamata-Viehweg vanishing theorem that we get the desired vanishings $H^i(X,\CO_X(P))=0$ for $i>0$ if $P$ is nef and $P^3>0$.

\smallskip\noindent
\textbf{Proof}. Follows immediately from \cite[Cor.\ 1.4.41]{Lazarsfeld:2004pag}.  Indeed, the proof shows that $\dim H^i(X,\CO_X(mP))=O(m^{3-i})$.

\smallskip
If $P$ is an effective divisor, nef already implies semi-ample.  Hence nef and semi-ample are equivalent conditions on effective divisors:

\begin{thm} \cite{OguisoK}\label{thm:sa}
If $P$ is effective (or more generally if $\kappa(X,P)\ge0$) and nef, then $P$ is semi-ample. 
\end{thm}

\begin{cor}\label{cor:smooth}
If $P$ is nef and $|nP|$ contains an effective surface for any $n\ge1$, then $|nmP|$ contains a smooth surface for some $m\ge1$.
\end{cor}

We can be more precise for ample divisors.
\begin{thm}
	\textbf{(Oguiso-Peternell Theorem \cite{OP} )}\label{Opt}
\end{thm}

\textit{Let $P$ be an ample divisor in a Calabi-Yau threefold.  Then}
\begin{enumerate}
	\item $|mP|$ \textit{is base point free for }$m \geq 5$
		\item $mP$ \textit{is very ample for }$m \geq 10$
\end{enumerate}
By Bertini's Theorem, we see that we can always find a smooth surface in $|5P|$.

\smallskip
We are primarily interested in smooth irreducible surfaces $P\subset X$.  Note that since $X$ is assumed projective, we have that $S$ is automatically projective, stronger than merely K\"ahler. We will see presently that if $P$ is a regular surface,  we get the desired vanishings $H^i(X,\CO_X(P))=0$ for $i>0$ without any additional hypotheses on the linear system $|P|$.
The reason is that the  numerical invariants of the surface $P$ are related to the properties of $P$ as a divisor in $X$.

\begin{prop}\label{prop:surfinv}
For $X$ and $P$ as above, we have
\begin{enumerate}
\item $\dim H^0(X,\CO(P))=p_g+1$ \label{eq:h0}
\item $\dim H^1(X,\CO(P))=q$\label{eq:h1}
\item $H^k(X,\CO(P))=0$ for\  $k\ge2$.\label{eq:h2}
\end{enumerate}
\end{prop}

\begin{cor}\label{cor:reg}
If the surface $P$ is regular, then $H^i(X,\CO_X(P))=0$ for $i>0$ and $\dim|P|=p_g$.
\end{cor}

\noindent
\textbf{Proof}.
We consider the short exact sequence 
\begin{equation}
0\to \CO_X\to \CO_X(P)\to \CO_X(P)|_P\to 0.  
\end{equation}
By the adjunction formula and the Calabi-Yau condition, we have $\CO_X(P)|_P\simeq K_P$. Using $H^1(X,\CO_X)=H^2(X,\CO_X)=0$ which is part of the Calabi-Yau condition, the associated long exact sequence of cohomology splits up into a short exact sequence,
\begin{equation}\label{eq:h0seq}
0\to H^0(X,\CO_X)\to H^0(X,\CO_X(P))\to H^0(P,K_P)\to 0,
\end{equation}
an isomorphism $H^1(X,\CO(P))\simeq H^1(P,K_P)$, and an exact sequence
\begin{equation}\label{eq:h2seq}
0\to H^2(X,\CO_X(P))\to H^2(P,K_P) \to H^3(X,\CO_X)\to H^3(X,\CO_X(P))\to 0.
\end{equation}
Taking dimensions in (\ref{eq:h0seq}) gives  \ref{eq:h0}.  By Serre duality on $P$, we have $H^1(P,K_P)\simeq H^1(P,\CO_P)^*$, which has dimension $h^{0,1}(P)=h^{1,0}(P)=q$.  So \ref{eq:h1} follows immediately from the isomorphism between (\ref{eq:h0seq}) and (\ref{eq:h2seq}).  By Serre duality on $X$ we get $H^3(X,\CO_X(P))\simeq H^0(X,\CO_X(-P))^*=0$, using the Calabi-Yau condition $K_X\simeq \CO_X$.  Since $H^2(P,K_P)\simeq H^{2,2}(P)$ and $H^3(X,\CO_X)$ are each 1-dimensional, (\ref{eq:h2seq}) implies that $H^2(X,\CO_X(P))=0$.  We trivially have $H^k(X,\CO(P))=0$ for $k>3$ for dimension reasons.  This completes the proof of \ref{eq:h2} and of the proposition.

\smallskip
Continuing to assume that $P$ is a smooth surface, note that if in addition $P$ is either ample, or more generally nef with $P^3>0$, then Kodaira vanishing or Kawamata-Viehweg vanishing implies that $H^1(X,\CO(P))=0$, so that $P$ is regular by Proposition~\ref{prop:surfinv}.  
 
\smallskip
If $P$ is merely semi-ample, then $|mP|$ is base point free for $m\gg0$, hence its restriction to $P$ is still basepoint free.  Since $P$ restricts to $K_P$ on $P$, we see that $|mK_P|$ is base point free and hence $\kappa(P)\ge 0$.  Furthermore $P$ cannot contain any  $(-1)$ curve $C$, since $K_P\cdot C=-1$ on $P$ is equivalent to $P\cdot C=-1$ on $X$, contradicting the fact that $P$ is nef.

Furthermore, if $P$ is ample, then the restriction $K_P$ of $\CO_X(P)$ to $P$ is still ample, hence $mK_P$ is very ample for $m>>0$ and $\kappa(P)=2$.  If $P$  nef with $P^3>0$, then $K_P^2=P^3>0$  and $P$ is a minimal surface with $\kappa\ge0$ as we just saw.  As a consequence of the Kodaira classification of minimal surfaces, we see that $K_P^2=0$ for minimal surfaces $P$ with $\kappa=0$ or 1.  It follows that $\kappa=2$ in this case as well.

Summarizing, we have proven

\begin{prop}\label{prop:kodairaprop}
Suppose that $P$ is a smooth surface which is also semi-ample as a divisor in $X$.  Then $P$ is a minimal surface with $\kappa\ge0$.  If in addition $P$ is ample, or more generally nef with $P^3>0$, then $P$ is a regular surface of general type.  
\end{prop}
Of course, if $|P|$ is very ample, then a general surface in $|P|$ is automatically smooth by Bertini's theorem.

\smallskip
Regarding $c_2$ we have

\begin{thm}\label{thm:miyaoka}
If $P$ is nef, then $c_2\cdot P\ge0$.
\end{thm}
This follows from \cite[Theorem 1.1]{miyaoka1987}.

\begin{cor}\label{cor:eff}
If $[P]$ is an ample class, or more generally if $[P]$ is nef with $P^3>0$, then it has an effective representative $P$.
\end{cor}

\noindent
\textbf{Proof}.  By Kodaira vanishing in the ample case, or Proposition~\ref{prop:big} and Kawamata-Viehweg in the more general case, we get $H^i(X,\CO(P))=0$ for $i>0$.  Then we have $h^0(\CO(P))=\chi(\CO(P)) = P^3/6+c_2.P/12 > 0$, the first equality coming from the vanishing of higher cohomology.

\begin{thm}
\cite{lazi2016morrisonkawamata} 	\label{Semi}
If $P$ is nef and $c_2(X)\cdot P>0$, then $P$ is semi-ample. \end{thm}

\begin{thm}
\textbf{(Lefschetz hyperplane theorem) }\label{Lht}

Let $P$ be an 
effective  ample 
divisor on a smooth projective variety $X$ of dimension $n$. 
Then the restriction map
$r_i : H^i(X, \IZ) \to H^i(P, \IZ)$
is an isomorphism for $i\leq n-2$ and injective for $i=n-1$. 
 \end{thm}
In particular, if $X$ is a Calabi-Yau threefold, then $\dim H^1(P, \IZ)=0$ so that $P$ is regular, and $\dim H^2(X, \IZ) \le \dim(H^2(P, \IZ))$.

\smallskip

\begin{thm}
(\textbf{Hodge index theorem)}\label{HI}
	
Assume $P$ is a compact surface then the cupproduct form on $H^2(P,\IR)$, restricted to $H^{1,1}_{\IR}(P)$, is non-degenerate and of signature $(1,h^{1,1}-1)$
\end{thm}

\begin{thm}
	\textbf{(Noether bound) }\label{Nt}
	
		Let $P$ be a smooth minimal surface of general type ($\kappa=2$). Then
\begin{eqnarray}
\frac{1}{2}K^2_P\geq p_g(P)-2
\end{eqnarray}
\end{thm}
In the case that $P$ is a smooth surface with $\kappa=2$, then $p_g=\frac{1}{6}P^3+\frac{1}{12}P.c_2(X)-1$ by Hirzebruch-Riemann-Roch and Proposition~\ref{prop:surfinv}. Since $K_P^2=P^3$, we conclude that $4P^3\geq P\cdot c_2(X)-36$ if $P^3$ is even or $4P^3\geq P\cdot c_2(X)-30$ if $P^3$ is odd.
In addition, all minimal smooth surfaces with $\kappa=0$ have $c_2(P)\leq 36$, which can be found in \cite{MR749574} and hence satisfy the same inequality. This implies that smooth ample or semi-ample divisors that correspond to smooth surfaces with $\kappa=0,2$ satisfy  $4P^3\geq P\cdot c_2(X)-C$, where $C=36$ when $P^3$ is even and $C=30$ when $P^3$ is odd.

\subsection{Proofs of Inequalities} \label{subsec:proofs}

 \vspace{0.2 cm}\begin{thm}
 	\textbf{(Inequality 4, Table \ref{tb:conditions}) }\label{I5}
 \end{thm}
 \textit{Let P be a 
smooth ample 
divisor inside the Calabi-Yau threefold $X$.  Then} \begin{equation}\label{eq:inequality5}
  h^{1,1}(X) \le P^3 + P\cdot c_2(X) -2
 \end{equation} 
 
\noindent
\textbf{Proof}. Since $P$ is 
smooth and ample 
in $X$,  the Lefschetz hyperplane theorem applies.  Therefore the restriction map $r: H^2(X,\IZ)\rightarrow H^2(P,\IZ)$ is an injection. Hence, 
\begin{equation}\label{lefineq}
dim(H^2(X,\IC))\leq dim(H^2(P,\IC)).
\end{equation}
 By the Hodge Decomposition we know that $dim(H^2(X,\IC)) = h^{1,1}(X)$ since $h^{2,0}(X)=0$ and $dim(H^2(P,\IZ)) = h^{1,1}(P)+2h^{2,0}(P)=h^{1,1}(P)+2p_g$.
 In addition, since $P$ is a regular surface,  the topological Euler characteristic of $P$ is  given by  $\chi(P)  = 2+b_2 =2+ 2p_g+h^{1,1}(P)$, while from (\ref{eq:chiX}) we also know that $\chi(P)  = P^3 + P\cdot c_2(X)$.
Hence, $h^{1,1}(P)=P^3 + P\cdot c_2(X)-2p_g-2$ which implies that (\ref{lefineq}) becomes $h^{1,1}(X)\leq h^{1,1}(P)+2p_g=P^3 + P\cdot c_2(X)-2$.

\begin{thm}
	\textbf{(Inequality 5, Table \ref{tb:conditions}) }\label{I6}
\end{thm}

\noindent
\textit{Let P be a smooth, 
big and nef  divisor inside the Calabi-Yau threefold $X$. Then the number of rational $(-2)$ curves on $P$ is bounded by } \begin{equation}
N_{-2} \le \frac{1}{6}(4P^3 +5 P\cdot c_2(X) )-1.
\end{equation} 
\textbf{Proof} A smooth 
nef and big 
divisor inside the Calabi-Yau threefold is a minimal surface of general type.
A consequence of the Hodge Index Theorem is that the number $N_{-2}$ of rational $-2$ curves in a surface of general type $P$ is bounded by
$N_{-2} \le \rho(P) - 1 $,where $\rho(P)$ is the Picard number of $P$, the rank of the group of divisor classes. This claim can be found in [\cite{MR749574},Prop.VII(2.5)]

In addition, the Picard number $\rho(P)$ is clearly bounded above by $h^{1,1}(P)$, as the Picard lattice of cohomology classes of divisors is a sublattice of $H^{1,1}(P,\IC)\cap H^2(P,\IZ)$.  Therefore $N_{-2} \le h^{1,1}(P)- 1 $. In the proof of  (\ref{eq:inequality5}), we saw that $h^{1,1}(P)=P^3 + P\cdot c_2(X)-2p_g-2$.  But  $\chi(\CO(P))=p_g+1$ by Proposition~\ref{prop:surfinv} and $\chi(\CO(P))=(2P^3+c_2\cdot P)/12$ by Hirzebruch-Riemann-Roch.  Combining these formulas, we conclude that $N_{-2} \le \frac{1}{6}(4P^3 +5 P\cdot c_2(X) )-1$.

\section{Examples}\label{app:examples}
In this section of the appendix, we collect examples supporting the discussion in the main text.  We begin with an example of an ample divisor class with no smooth representative.  We then follow with examples of  $SU(2)$ and $SU(3)$ gauge theories which Higgs to the quintic.

\subsection{An ample divisor class with no smooth representative}\label{app:singular}
Referring to \cite{Candelas:1994tpt}, we let $X$ be a smooth Weierstrass elliptic fibration over $\IP^2$, equivalently the blowup of a weighted hypersurface $\hat{X}$ of degree~18 in $\IP(1,1,1,6,9)$.  The closure $\overline{\mathcal{K}(X)}$ of the K\"ahler cone is generated by two classes, denoted by $H$ and $L$.  The dual Mori cone generators are denoted by $h$ and $\ell$. Each of the classes $H$ and $L$ are nef but not ample (we have $H\cdot\ell=0$ and $L\cdot h=0$).  But $H+L$ is in the interior of the nef cone hence is ample (cf.\ the discussion following Theorem~\ref{Nef}).  We study the surfaces in $|H+L|$ and show that all are singular. 

The blowup of $\hat{X}$ is performed along the singular locus $x_1=x_2=x_3=0$ (a single point in $\hat{X}$ due to the imposition of the defining weight~18 equation), with exceptional divisor $E\simeq\IP^2$.  The blowup guarantees that the projection to the first three coordinates gives a well-defined map $X\to \IP^2$ with elliptic fibers, the base being embedded in $X$ as the section $E$.  The Mori generator $\ell$ is a line in $E\simeq \IP^2$ and the Mori generator $h$ is the class of the elliptic fiber.  The divisor class $L$ is the pullback of $\CO_{\IP^2}(1)$ to $X$, and in particular is represented by the proper transforms of any of the surfaces defined by $x_i=0$, $i=1,2,3$.   The class $H$ is defined as $3L+E$.  In particular, $H$ projects to a class of weight~3 in $\IP(1,1,1,6,9)$.  Thus $2H$ projects to a class of weight 6 and $3H$ projects to a class of weight 9.  Furthermore, it can be checked that
the proper transform of $x_4=0$ is in the class $2H$ and the proper transform of $x_5=0$ is in the class $3H$. 

We now examine the class $H+L=4L+E$, which has weight 4 after projection to $\IP(1,1,1,6,9)$.  But the only weight~4 polynomials in $\IP(1,1,1,6,9)$ are just the degree 4 homogeneous polynomials $f(x_1,x_2,x_3)$ in the homogeneous coordinates of the base $\IP^2$.  The proper transform of $f=0$ is in the class $4L$.  Thus any effective divisor $D$ in $|H+L|$ contains as a component a surface in $|4L|$,
which is simply the restriction $S$  of the elliptic fibration to a plane curve $C$ in the base of degree 4.  We conclude that $D=S\cup E$, which is singular along $S\cap E$.  This last is just the curve
$C$ identified as a curve in the section $E$.

\subsection{$SU(2)$}\label{app:SU2}
In our first example, an $SU(2)$ gauge theory, the geometry is a singular quintic with an $A_1$ singularity along a line $L$, and smooth otherwise.  For definiteness, we choose homogeneous coordinates $(x_0,\ldots,x_4)$ on $\IP^4$ so that $L$ is defined by $x_0=x_1=x_2=0$.  Then the equation of the quintic has the form
\begin{equation}
\sum_{i,j=0}^2x_ix_jf_{ij}(x_3,x_4)=0,
\label{eq:singline}
\end{equation}
where the $f_{ij}$ are homogeneous polynomials of degree~3.

More generally, we can find a quintic with an $SU(2)$ on any curve $C$ which can be defined by the simultaneous vanishing of a collection of homogeneous polynomials $q_i(x)=q_i(x_0,\ldots,x_4)$ of degrees $d_i\le2$.  In addition to the case of the line above $\{d_i\}=\{1,1,1\}$, we will also consider the cases where $C$ is a plane conic $\{d_i\}=\{1,1,2\}$ or a twisted cubic $\{d_i\}=\{1,2,2,2\}$.  In general, letting $f_{ij}(x)=f_{ij}(x_0,\ldots x_4)$ denote generic  homogeneous polynomials of degrees $5-d_i-d_j$, the quintic defined by the equation
\begin{equation}
\sum_{i,j}q_i(x)q_j(x)f_{ij}(x)=0
\label{eq:singC}
\end{equation}
has an $A_1$ singularity at the generic point of $C$.  The assumption $d_i\le2$ is needed to ensure that $5-d_i-d_j>0$ and so nonvanishing $f_{ij}(x)$ exist.

Note that we are not assuming that the $q_i$ are independent (as in the case of the line above), so there could be more than one way to choose the $f_{ij}(x)$ to get a fixed quintic.  The twisted cubic is an example where such an ambiguity arises, with a linear syzygy relating the three quadratic terms.

Returning to the case of an $SU(2)$ on a line $L$, we now identify the matter.  
At a point $(0,0,0,x_3,x_4)\in L$ (which hereafter we simply write as $(x_3,x_4)\in L$), the type of the transverse singularity can be identified by the matrix
\begin{equation}
A(x_3,x_4)=\left(
\begin{array}{ccc}
f_{00}(x_3,x_4)&f_{01}(x_3,x_4)&f_{02}(x_3,x_4)\\
f_{10}(x_3,x_4)&f_{11}(x_3,x_4)&f_{12}(x_3,x_4)\\
f_{20}(x_3,x_4)&f_{21}(x_3,x_4)&f_{22}(x_3,x_4)
\end{array}
\right).
\end{equation}
We have a transverse $A_1$ singularity at $(x_3,x_4)$ when $\det A(x_3,x_4)\ne0$.  We assume that the $f_{ij}$ are chosen generically, so that $\det A(x_3,x_4)$ is a degree~9 homogeneous polynomial vanishing at 9~distinct points, which are generically $A_2$ singularities.  The $SU(2)$ gauge theory therefore has $N_f=9$, with the matter localized at the zeros of $\det A$.    Similar methods can be used to locate the matter starting from equations of the form (\ref{eq:singC}).  However, in this paper we primarily concerned with the value of $N_f$ rather than the more precise information of the location of the matter.   Later in this example, we will compute $N_f=9$ by a different method which will generalize in a straightforward manner.

Blowing up the singular quintic gives a Calabi-Yau $X$ with $h^{1,1}(X)=2$.  The cohomology generators are $H$, the proper transform of the hyperplane class of the quintic, and $E$, the exceptional divisor of the blowup.  We need to compute the triple intersection numbers of $H$ and $E$, their intersections with $c_2=c_2(X)$, and the generators of the K\"ahler cone.

We proceed by first blowing up $L$ inside $\IP^4$ to obtain the blown-up fourfold $\widetilde{\IP^4}$.  Then $H^{1,1}(\widetilde{\IP^4})$ is generated by $\mathbf{H}$, the proper transform of the hyperplane class of $\IP^4$, and $\mathbf{E}$, the exceptional divisor.  We have
\begin{equation}
H=\mathbf{H}|_X,\qquad E=\mathbf{E}|_X.
\label{eq:resttox}
\end{equation}
Since $X$ is obtained by blowing up a quintic (degree~5) with a multiplicity~2 singularity along $L$, we get for the class $[X]\in H^{1,1}(\widetilde{\IP^4})$ of $X$
\begin{equation}
[X]=5\mathbf{H}-2\mathbf{E}.
\label{eq:x}
\end{equation}

To compute the triple intersections on $X$, we lift to classes to $\widetilde{\IP^4}$ using (\ref{eq:resttox}) and then restrict the corresponding triple intersection on $\widetilde{\IP^4}$ to $X$.  Using~(\ref{eq:x}) we get
\begin{equation}\begin{split}
H^3=\mathbf{H}^3\left(5\mathbf{H}-2\mathbf{E}\right),\ H^2E=\mathbf{H}^2\mathbf{E}\left(5\mathbf{H}-2\mathbf{E}\right),\\ HE^2=\mathbf{H}\mathbf{E}^2\left(5\mathbf{H}-2\mathbf{E}\right),\ E^3=\mathbf{E^3}\left(5\mathbf{H}-2\mathbf{E}\right).
\end{split}
\label{eq:tripinta}
\end{equation} 
To finish the calculation, we just need the four-fold intersection products $\mathbf{H}^i\mathbf{E}^{4-i}$ on $\widetilde{\IP^4}$.  This is a standard calculation in algebraic geometry, using
Segre classes~\cite{Fulton:1984it}.  The Segre class $s(S,M)$ of a submanifold $S\subset M$ is the inverse of the total chern class of the normal bundle $N_{S,M}$ of $S$ in $M$:
\begin{equation}
s(S,M)=c(N_{S,M})^{-1},
\end{equation}
a cohomology class on $S$.    Now suppose that we have a birational mapping of manifolds $f:M\to N$ with $T=f(S)$ also a manifold.  Then we have
\begin{equation}
f_*\left(s(S,M)\right)=s(T,N),
\end{equation}
i.e.\ Segre classes are invariant under birational pushforward~\cite[P.\ 76]{Fulton:1984it}.

In the special case where $S$ is a divisor, we have $N_{S,M}$ is a line bundle, and $c_1(N_{S,M})$ is the restriction of the cohomology class of $S$ itself to $S$.
So $c(N_{S,M})$ is the restriction of $1+S$ to $S$.  Specialing $S\subset M$ to $\mathbf{E}\subset \widetilde{\IP^4}$ and inverting, we get
\begin{equation}
s(\mathbf{E},\widetilde{\IP^4}) = \mathbf{E} - \mathbf{E}^2 +\mathbf{E}^3 - \mathbf{E}^4.
\end{equation}
For the projection $\pi:\widetilde{\IP^4}\to \IP^4$ we have $\pi(E)=L$.  Since $c_1(N_{L,\IP^4})=3p$ ($p$ being the class of a point), we invert $c(N_{L,\IP^4})=L+3p$ on $L$ and get
\begin{equation}
s(L,\IP^4)=L-3p.
\end{equation}
Then $\pi_*(\mathbf{E} - \mathbf{E}^2 +\mathbf{E}^3 - \mathbf{E}^4)=L-3p$ gives
\begin{equation}
\pi_*(\mathbf{E})=0,\ \pi_*(\mathbf{E}^2)=0, \pi_*(\mathbf{E}^3)=L, \pi_*(\mathbf{E}^4)=3p.
\label{eq:epowers}
\end{equation}
Letting $h$ be the hyperplane class of $\IP^4$ with $h^4=1$, and $H=\pi^*h$, we can now compute the four-fold intersections on $\widetilde{\IP^4}$ by
\begin{equation}
\mathbf{H}^i\mathbf{E}^{4-i}=\pi_*(\mathbf{H}^i\mathbf{E}^{4-i})=\pi_*((\pi^*h^i)\mathbf{E}^{4-i})=h^i\pi_*(\mathbf{E}^{4-i}).
\end{equation}
Combining with (\ref{eq:epowers}) we get
\begin{equation}\begin{split}
\mathbf{H}^4=h^4=1,\ \mathbf{H}^3\mathbf{E}=h^3\pi_*(\mathbf{E})=0,\ \mathbf{H}^2\mathbf{E}^2=h^2\pi_*(\mathbf{E^2})=0,\\
\mathbf{H}\mathbf{E}^3=h\pi_*(\mathbf{E}^3)=hL=1,\ \mathbf{E}^4=3.
\end{split}
\end{equation}
Plugging these into (\ref{eq:tripinta}) we get
\begin{equation}\label{eq:triple-intersections-SU2-9}
H^3=5,\ H^2E=0,\ HE^2=-2,\ E^3=-1.
\end{equation}
Since $E$ is a ruled surface over $L\simeq\IP^1$, it is the blowup of a Hirzebruch surface at $N_f$ points.  However, $E^3$ is the self-intersection of the canonical bundle of $E$, which is $8-N_f$.  So $E^3=-1$ is equivalent to $N_f=9$.

For a smooth surface $S$ on any Calabi-Yau we have $S^3+S\cdot c_2=c_2(S)$.  Applying this to $H$, a quintic surface in $\IP^3$ with $c_2=55$ we get $H\cdot c_2=50$.  From the description of $E$ as the blowup of a Hirzebruch surface at 9 points we get $c_2(E)=13$, as a Hirzebruch surface has $c_2=4$ and each blowup adds 9.  Combining with $E^3=-1$ we get $E\cdot c_2=14$.  Summarizing:
\begin{equation}
H\cdot c_2=50,\qquad E\cdot c_2=14.
\end{equation}

Finally, we turn to the K\"ahler cone, which is most easily computed from the dual Mori cone. The calculation is elementary albeit a bit lengthy.  We provide all of the details in this case to illustrate the ideas.  In the other examples in this and the following section, we omit details in the calculation of the Mori cone.   In some cases, we do not have a mathematical proof that we have found all of the Mori generators, but we provide justification by checking consistency with physics.

We coordinatize the Mori cone by identifying the class $[D]$ of a curve $D\subset X$ with the ordered pair 
\begin{equation}\label{eq:coords}
(D\cdot H,D\cdot E)\in\IZ^2.
\end{equation}  
Alternatively, if desired we could identify of pair of curve classes which generate $H_2(X,\IZ)$ and express all curve classes in terms of the two chosen generators.  While that approach might clarify the geometry, using our coordinates is simpler.

We identify irreducible curves $D\subset X$ with help of the blowdown map $\pi:X\to Y$ which contacts $E$ to the
line $L$ in the singular quintic $Y\subset\IP^4$.  The restriction of $\pi$ to $E$ exhibits $E$ as a ruled surface over $L$.  Let $r$ be the class of the generic fiber.  Since a general hyperplane in  $Y$ intersects $L$ at one point, its proper transform $H$ in $X$ is disjoint from the fiber $r$ over any other point of $L$.  Thus $H\cdot r=0$.  Furthermore, $E\cdot r=-2$ because the curve $r$ can be viewed as the exceptional curve of a transverse $A_1$ singularity.  Thus $r$ has coordinates $(0,-2)$.

There are $N_f=9$ special fibers which split into a pair of $\IP^1$'s.  Since each $\IP^1$ in this pair is orthogonal to $H$, the two classes lie in the same 1-dimensional subspace of the two-dimensional  $H_2(X,\IZ)$ and are therefore proportional.  We conclude that each of these $\IP^1$'s has class $r/2$ and coordinates $(0,-1)$.

If $D$ is not contained in a fiber of $\pi$, then $\pi(D)$ is a curve in $Y\subset \IP^4$ of some degree $d>0$.  For example, $\pi(D)$ can be a line, $d=1$. 
There are two cases to consider separately: $\pi(D)=L$ or $\pi(D)\ne L$. 

We consider the latter case first.  Let $d$ be the degree of $\pi(D)$ as a curve in $X\subset\IP^4$, so that $H\cdot D=d$.  Since $\pi(D)\ne L$, the curves $\pi(D)$ and $L$ meet at finitely many points (possibly none).  Equivalently, $D$ and $E$ meet at finitely many points.  Putting $k=E\cdot D\ge 0$, we conclude that the coordinates of $D$ are $(d,k)$.

We now show that $k\le d$.
Choose a hyperplane $P\subset \IP^4$ containing $L$ 
but not containing $\pi(D)$.  Then $P$ meets $\pi(D)$ in $d$ points (including multiplicity) by the definition of degree.  On the other hand, $P$ meets $\pi(D)$ in at least $k$ points (including multiplicity), namely those contained in $L$.  Thus $k\le d$, as claimed.

We now exhibit a curve $D$ with $d=k=3$.  Choose a two-plane $Q\subset \IP^4$ containing $L$.  Since $Y\subset \IP^4$ is a quintic, we have that $Q\cap Y$ is a degree~5 plane curve, including multiplicities.  However $L\subset Q\cap Y$, and $L$ occurs with multiplicity 2 in $Q\cap Y$ due to the $A_1$ singularity.  It follows that
\begin{equation}\label{residual}
Q\cap Y=2L+D
\end{equation}
for some degree 3 curve $D$, i.e.\ $d=3$ for the curve $D$.  Since $D$ and $L$ are contained in the same plane $Q$, they meet in exactly 3 points, and $k=3$ as claimed.  Thus $D$ has coordinates $(3,3)$.  

Thus the Mori cone is spanned (over $\IQ$) by the curve classes with coordinates $(0,-1)$, $(1,1)$, and the curves $D$ with $\pi(D)=L$.

We are now ready to turn to the case $\pi(D)=L$, i.e.\ curves $D\subset E$, and show that these classes are already in the span of the curve classes found above.  

Since the half-fibers have self-intersection $-1$, we can blow down either of the $\IP^1$s in the 9 singular fibers and get a $\IP^1$-bundle over $\IP^1$.  Thus $E$ is the blowup of a Hirzebruch surface.  Note that the Mori cone of a blown up Hirzebruch surface is generated by the exceptional curves and some sections, as discussed for example in \cite{Jefferson:2018irk}.  The exceptional curves which are not sections are among the half-fibers $r/2$ which we have already accounted for in the Mori cone.

To determine the possible coordinates of the sections, we describe $E$ as a hypersurface inside $\IP^1\times\IP^2$ by viewing $(x_0,x_1,x_2)$ as homogeneous coordinates for $\IP^2$ and $(x_3,x_4)$ as homogeneous coordinates for $\IP^1$ in (\ref{eq:singline}).  Thus $E$ is a hypersurface of bidegree $(3,2)$.  

Now a section of $E$ can be thought of as the image of a map $\IP^1\to E\subset \IP^1\times\IP^2$.  Thus any section $D$ of $E$ is the graph of a map $\IP^1\to \IP^2$.   Let $s$ be the degree of this map.  We now compute the coordinates of $D$ in terms of $s$.

First, we have $H\cdot D=1$, since $H$ meets $E$ in a fiber of $E$, which in turn meets $D$ in exactly one point since $D$ is a section.

Next, $E\cdot D$ is equal to the degree of the restriction $(K_E)|_D$ to $D$ of the canonical bundle  of $E$.  Since the canonical bundle of $\IP^1\times \IP^2$ is $\CO(-2,-3)$ and $E$ is a section of $\CO(3,2)$, the adjunction formula tells us that $K_E$ is  $\CO(1,-1)$, which has degree $1-s$ after restricting to $D$.  Thus the section $D$ has coordinates $(1,1-s)$.  
Since these classes are all in the cone spanned by the curve classes with coordinates $(0,-1)$ and $(1,1)$, we see that
the Mori cone is spanned by  the curve classes with coordinates 
\begin{equation}\label{eq:morigen}
(0,-1),\  (1,1).
\end{equation}

Finally, the K\"ahler cone is generated by
\begin{equation}
H,\ H-E,
\end{equation}
the dual basis to (\ref{eq:morigen}).

\smallskip
Before turning to other examples, we first make some general observations.  Suppose we have a Calabi-Yau threefold $Y$ with a generic $A_1$ singularity along a smooth curve $C$ of genus $g$, enhancing to $A_2$ at $N_f$ distinct points.  Let $\pi:X\to Y$ be the blowup of $C$, with exceptional divisor $E$.  Then $E$ is a ruled surface over $C$ with generic fiber $r$, and $N_f$ special fibers consisting of pairs of $\IP^1$'s, each of class $r/2$.   Let $\{D_i\}$ be any collection of divisors in $H^2(Y,\IZ)$, and we continue to denote their pullbacks to $X$ by the same symbols.  Then by similar methods to the above example, we compute
\begin{equation}
D_i D_jE=0,\ D_i E^2= -2D_i C, E^3=8-8g-N_f,
\label{eq:a1trip}
\end{equation}
while the triple intersections of the $D_i$ are identical when computed on either $X$ or $Y$.  The intersection $D_iC$ is computed on $Y$, while the triple intersections in (\ref{eq:a1trip}) are computed on $X$.

We have done computations for quintics with $SU(2)$ on various curves $C$.   Suppose that $C\subset Y\subset\IP^4$ has degree $c$ and genus $g$.  Since $c_1(N_{C,\IP^4})=5d+2g-2$, 
we compute $c(N_{C,\IP^4})=[C]+(5d+2g-2)p$ so that
\begin{equation}
s(C,\IP^4)=C+(5d+2g-2)p.
\end{equation}
Birational invariance of Segre classes then reads
$\pi_*(\mathbf{E} - \mathbf{E}^2 +\mathbf{E}^3 - \mathbf{E}^4)=C-(5d+2g-2)p$, giving
\begin{equation}
\pi_*(\mathbf{E})=0,\ \pi_*(\mathbf{E}^2)=0, \pi_*(\mathbf{E}^3)=C, \pi_*(\mathbf{E}^4)=(5d+2g-2)p,
\label{eq:epowersgen}
\end{equation}
hence
\begin{equation}
\mathbf{H}^4=1,\ \mathbf{H}^3\mathbf{E}=0,\ \mathbf{H}^2\mathbf{E}^2=0,\ 
\mathbf{H}\mathbf{E}^3=hC=d,\ \mathbf{E}^4=5d+2g-2.
\end{equation}
It follows that
\begin{equation}
H^3=5,\ H^2E=0,\ HE^2=-2d,\ E^3=4-4g-5d.
\label{eq:tripintc}
\end{equation}
Since a $\IP^1$-bundle over $C$ has $K^2=8-8g$, the surface $E$ must be a $\IP^1$-bundle over $C$ blown up at $N_f=(8-8g)-(4-4g-5d)=5d+4-4g$ points.  

Combining $c_2(E)=4-4g+N_f=5d+8-8g$ (since a $\IP^1$-bundle over a curve of genus $g$ has $c_2=2(2-2g)=4-4g$ and each blowup adds 1) with $c_2(E)=E^3+Ec_2$, we get $Ec_2=10d+4-4g$.

We collect the results in the following table.  Here $d'$ is the degree of $C\subset\IP^4$.

\begin{equation}
\begin{array}{|c|c|c|c|c|c|c|c|c|c|c|}\hline
d'&g&{\rm Mori\ gens}&{\rm Kahler\ gens}&N_f&H^3&H^2E&HE^2&E^3&c_2H&c_2E\\ \hline
1&0&(1,1),\ (0,-1)&H,\ H-E&9&5&0&-2&-1&50& 14\\ \hline
2&0&(1,2),\ (0,-1)&H,2H-E&14&5&0&-4&-6&50&24 \\ \hline
3&0&(1,2),\ (0,-1)&H,\ 2H-E&19&5&0&-6&-11&50&34 \\ \hline

\end{array}
\label{eq:su2results}
\end{equation}
The only information in (\ref{eq:su2results}) which does not follow immediately from (\ref{eq:tripintc}) and the following paragraphs are the Mori generators and the K\"ahler generators.  We have continued to coordinatize the Mori cone by  $[D]\mapsto (H\cdot D,E\cdot D)$.  The K\"ahler generators are immediately deduced from the Mori generators by duality, 
so we need only describe the Mori generators.

The method is a straightforward adaptation of the case of a line.  We consider the blowdown $\pi:X\to Y$ to a quintic $Y$ with an $A_1$ singularity along $C$, and separately consider the
cases $\pi(D)=C$ and $\pi(D)\ne C$.   We have half-fibers $r/2$ with coordinates $(0,-1)$.  If $\pi(D)\ne C$, then $[D]$ has coordinates $(d,k)$, with $d>0$ and $k\ge0$ exactly as in the case $C=L$.  We find a curve $D$ which maximizes the slope of the ray from the origin through $(d,k)$ and as before, we can show that the case $\pi(D)=C$ does not produce any new classes.  Then the Mori cone is generated by $(d,k)$ and $(0,-1)$.

In the case $d=2,g=0$, such curves are well-known to be contained in a unique two-plane $Q$~\cite[Example 6.4.2]{Hartshorne:1977ag}, which intersects $Y$ in a degree 5 curve $Y\cap Q$.  This intersection contains $C$ with multiplicity 2.  Considering degrees, we see that we must have
\begin{equation}
Y\cap Q = 2C+D
\end{equation}
for some line $D$, $d=1$.  Since $D$ meets $C$ in 2 points by plane geometry, we see that $k=2$ and $D$ has coordinates $(1,2)$ and slope 2.

We now show that any other irreducible curve $D$ with coordinates $(d,k)$ has slope $k/d<2$. It follows that Mori generators are those appearing in the second line of (\ref{eq:su2results}).

To see this, our previous argument shows that the line $D$ above is the \emph{only} curve in the quintic $Y$ other than $C$ which is contained in $Q$. Since any other curve $D'$ is not contained in $Q$, we can find a hyperplane $P$ containing $Q$ (hence containing $C$) which does not contain $D'$.  The same argument as in the case of an $SU(2)$ on a line shows that $k\le d$, and we are done since the slope of the rays associated to these curves are at most 1.

In the case where $C$ is a twisted cubic $d=3,g=0$, we found the curve $D$ whose coordinate ray has maximal slope experimentally by a computer search.  We simply describe this curve.

First, we note that the curve $C$ is defined by the vanishing of homogeneous polynomials $\ell_1,q_2,q_3,q_4$ of degrees~1,2,2,2. 
The hypersurface $\ell_1=0$ intersects the singular quintic threefold $Y$ in a quintic surface $S$ which is also singular along the curve $C$. We let $\tilde{S}$ be the proper transform of $S$ inside the Calabi-Yau $X$.  The surface $\tilde{S}$ is the blowup of $S$ along $C$, and is a smooth surface assuming that we have chosen the singular quintic $Y$ containing $C$ generically.

We now calculate intersections on $\tilde{S}$ by blowing up $S\subset \IP^3$ along $C$ using exactly the same method we previously used to find intersections on $X$ by blowing up $Y\subset\IP^4$ along $C$.  We state results without providing all of the supporting calculations.

We denote the exceptional divisor of the blowup $\widetilde{\IP^3}$ of $\IP^3$ by $\mathbf{F}$, the proper transform of a hyperplane by $\mathbf{H}$, the projection $\widetilde{\IP^3}\to\IP^3$ by $\pi$, the restriction of $\mathbf{F}$ to $\tilde{S}$ by $F$ and the restriction of $\mathbf{H}$ to $\tilde{S}$ by $H$.  Calculating Segre classes as in the $SU(2)$ cases, we get $\pi_*(\mathbf{F})=0,\ \pi_*(\mathbf{F^2})=-C,\ \mathbf{F}^3=-10$, which yields
\begin{equation}\label{intp3t}\mathbf{H}^3=1,\ \mathbf{H}^2\mathbf{F}=0,\ \mathbf{H}\mathbf{F}^2=-3,\ \mathbf{F}^3=-10.
\end{equation}
Since $\tilde{S}$ has class $5\mathbf{H}-2\mathbf{F}$ in $\widetilde{\IP^3}$, we calculate products of $H$ and $F$ in $\tilde{S}$ by replacing $H$ and $F$ by $\mathbf{H}$ and $\mathbf{F}$ respectively, multiplying by $5\mathbf{H}-2\mathbf{F}$, then calculating the resulting intersection on $\widetilde{\IP^3}$ using (\ref{intp3t}).  We obtain after calculation
\begin{equation}
F^2=5,\ HF=6,\ H^2=5.
\end{equation}
The desired curve $D$ has class $7H-4F$. For this class, we compute $H\cdot D = 7H^2-4HF=11$ and $F\cdot D=7HF-4F^2=22$, So $D$ has coordinates $(11,22)$, slope 2.  This ray is indicated (\ref{eq:su2results}).  To show that $D$ is in the Mori cone, we just have to show that there is an effective curve in this class.

By the adjunction formula for $\tilde{S}\subset\widetilde{\IP^3}$, the canonical class $K_{\tilde{S}}$ of $\tilde{S}$ is $((-4\mathbf{H}+\mathbf{F})+(5\mathbf{H}-2\mathbf{F}))|_{\tilde{S}}=H-F$.  Since $K_{\tilde{S}}$ has degree $H(H-F)=-1<0$, the class
$K_{\tilde{S}}$ is not effective, i.e.\ $p_g(\tilde{S})=0$ and hence $\chi(\CO_{\tilde{S}})=1$.  Riemann-Roch then gives $\chi(\CO(D))=(1/2)D(D-K_{\tilde{S}})+1=(1/2)(7H-4F)(6H-3F)+1=1>0$.
Also $H^2(\CO(D))$ is Serre dual to $H^0(K_{\tilde{S}}-D)=H^0(\CO(-6H+3F))$, which is zero since $-6H+3F$ has negative degree $H(-6H+3F)=-12$.  Thus $1=\dim H^0(\CO(D))-\dim H^1(\CO(D))$ and so $H^0(\CO(D))$ is nonzero.  Hence $D$ is effective, as claimed.

\subsection{$SU(3)$}\label{app:SU3}

To achieve an $SU(3)$ geometry on a line $L$, we inspect the $SU(2)$ geometry (\ref{eq:singline}) and see that by reinterpreting $(x_0,x_1,x_2)$ as homogeneous coordinates on $\IP^2$, the same equation describes $E$ as a ruled surface over the $\IP^1$ with homogeneous coordinates $(x_3,x_4)$.  The fibers are degree 2 curves in $\IP^2$ which are generically isomorphic to $\IP^1$, except over the $N_f=9$ points where $\det A(x_3,x_4)=0$.  For those points, the degree 2 curve factors into a product of linear terms and the fiber is a pair of lines, corresponding to the geometry of an $SU(2)$ enhancement.

This description immediately suggests a way to achieve an $SU(3)$: we require the degree~2 curve in every fiber to factor.  This can be achieved if the equation of $E$ factors as
\begin{equation}
\left(x_0g_0(x_3,x_4)+x_1g_1(x_3,x_4)+x_2g_2(x_3,x_4)
\right)
\left(x_0h_0(x_3,x_4)+x_1h_1(x_3,x_4)+x_2h_2(x_3,x_4)
\right).
\label{eq:factor}
\end{equation}
In (\ref{eq:factor}), the degrees of the polynomials $g_j$ and $h_j$ are fixed by an integer $0\le n\le 3$: the $g_j$ all have degree $n$ and the $h_j$ all have degree $3-n$.  By construction, each term in (\ref{eq:factor}) has degree~5 in the full set of variables $(x_0,\ldots,x_4)$ so is the equation of a quintic.  But this is not a good quintic when viewed as a hypersurface in $\IP^4$, since it visibly has two components, one of degree $n+1$ and the other of degree $4-n$.  This is easily fixed by adding terms of order greater than two in $(x_0,x_1,x_2)$
\begin{equation}
\left(x_0g_0+x_1g_1+x_2g_2
\right)
\left(x_0h_0+x_1h_1+x_2h_2
\right)+\ldots.
\label{eq:su3line}
\end{equation}
For generic $g_j,h_j$, and higher order terms, the quintic (\ref{eq:su3line}) has an $SU(3)$ geometry along a line, and no other singularities.

We will blow up this geometry twice to a smooth Calabi-Yau threefold which Higgses to the quintic.  The choice of blowup depends on the ordering of the factors in (\ref{eq:factor}).  For this reason, the construction is not symmetric in $g$ and $h$.  In particular,
switching $g$ and $h$, and replacing $n$ with $3-n$ gives the same geometry (\ref{eq:su3line}) but a different smooth Calabi-Yau.  These distinct Calabi-Yaus are related by a flop.  

We start by blowing up $\IP^4$ along the line $L$, just as we did in the $SU(2)$ case, and now consider the proper transform $Z$ of the quintic $Y$ defined by (\ref{eq:su3line}).  The exceptional divisor $E$ is still fibered over $L$, and by construction it splits into two components, each component being a $\IP^1$-bundle over $L$, i.e.\ a Hirzebruch surface. The two components intersect in a section $F$ of $E$ over $L$.  While it is clear that $Z$ is smooth away from $F$ and at the generic point of $F$, there is nothing to prevent $Z$ from having conifolds at finitely many points of $F$.  We will perform a blowup of $Z$ along one of the Hirzebruch surfaces which will both detect the conifolds and resolve them by small resolutions.  A choice of small resolution will be made in the process.

As a preliminary, we show how blowing up a surface in a threefold can detect a singularity in the threefold.  First, consider a smooth surface in a smooth threefold.  We can choose local analytic coordinates $(x,y,z)$ in the threefold so that the surface is defined by $z=0$.  Since there is only one equation, blowing up $z=0$ does nothing, and the proper transform of the surface is isomorphic to the surface being blown up.

By contrast, suppose a smooth surface passes through a conifold point.  We can choose local analytic coordinates $(w,x,y,z)$ so that the conifold is defined by $wx=yz$ and the surface is defined by $w=y=0$.  Now if we blow up $w=y=0$, we get two coordinate patches.  In the first patch we have a new coordinate $u=w/y$, leaving coordinates $(u,x,y,z)$ after eliminating $w$ via $w=uy$. Making this substitution into the equation of the conifold and factoring out $y$, we get $ux=z$, i.e.\ $z$ can be eliminated as well, leaving independent coordinates $(u,x,y)$, i.e.\ this patch of the blowup is a smooth threefold.  The blowdown map is seen to be
\begin{equation}
(u,x,y)\mapsto (w,x,y,z)=(uy,x,y,ux).
\label{eq:bd1}
\end{equation}
The inverse image of the conifold point in this patch is $\{(u,0,0)\}$, a copy of $\IC$.

We have a second coordinate patch described in terms of a new coordinate $v=y/w$.  A similar calculation gives coordinates $(v,w,z)$ and blowdown map
\begin{equation}
(v,w,z)\mapsto (w,x,y,z)=(w,vz,vw,z).
\label{eq:bd2}
\end{equation}
The inverse image of the conifold point in this patch is $\{(v,0,0)\}$, another copy of $\IC$.  Since the first coordinates in these two patches are related by $v=u^{-1}$,  we see that the conifold gets blown up to $\IP^1$, and we have a small resolution.
Furthermore, the local forms  (\ref{eq:bd1}),\ (\ref{eq:bd2}) of the blowdown map show that the exceptional $\IP^1$ is identified with the exceptional $\IP^1$ of the blowup of surface $w=z=0$ with coordinates $(x,y)$.  

Now the divisor $w=0$ in the singular threefold has two component divisors: $w=y=0$ and $w=z=0$.  Blowing up the first introduces an exceptional $\IP^1$ in its proper transform.   We now show that the blowup does not change the other divisor.  If we consider the divisor $w=z=0$ and make the coordinate change $w=uy$, recalling that the exceptional divisor is $y$, we get the proper transform of this divisor is $u=z=0$, or just $u=0$ since $z=uy$ as discussed above.  The inverse image via (\ref{eq:bd1}) of the conifold point inside $u=0$ is just $(0,0,0)$, so the proper transform of the divisor $w=z=0$ is isomorphic to the original divisor inside this coordinate patch.  A similar calculation in the other coordinate patch completes the verification of our assertion.

This gives us our strategy for identifying and resolving the conifolds: by blowing up one Hirzebruch surface, we introduce exceptional $\IP^1$'s in its proper transform without changing the other Hirzebruch surface.  We will see this explicitly in our $SU(3)$ model after further calculation.  This process involves a choice and is asymmetric, related by flops.  Furthermore, since we are using algebraic blowups, the resulting smooth threefold is guaranteed to be K\"ahler.

We now implement this strategy by blowing up the Hirzebruch surface $S$ corresponding to the first factor of (\ref{eq:factor}).   It can be shown that 
\begin{equation}
S\simeq\left\{
\begin{array}{cl}
\IF_0&n\ {\rm even}\\
\IF_1&n\ {\rm odd}
\end{array}
\right.
\end{equation}
but we do not need this, as $K_S$ (needed for Segre classes) can be computed by other techniques.   Instead, we note that the exceptional divisor $\IE$ of $\widetilde{\IP^4}$ is a trivial $\IP^2$-bundle over $L$, i.e.\ is isomorphic to $\IP^1\times\IP^2$,  essentially because the homogeneous coordinates $(x_0,x_1,x_2)$ on the fiber are independent of the coordinates on $L$.   The equation of the surface $S$ has degree $n$ in the $\IP^1$ variables and degree $1$ in 
the $\IP^2$ variables.  If we let $h_1\in H^2(\IP^1,\IZ)$ and $h_2\in H^2(\IP^2,\IZ)$ be the respective generators, we therefore have for the class of $S$
\begin{equation}
[S]=nh_1+h_2.
\end{equation}
Then we can generate $H^2(S,\IQ)$ by $f=h_1|_S$ and $h=h_2|_S$, where $f$ is a fiber of the Hirzebruch surface and $h$ is a section.   We compute
\begin{equation}
h^2=h_2^2\left(nh_1+h_2\right)=n.
\end{equation}
We also have
\begin{equation}
f^2=0,\ hf=1,
\end{equation}
either by lifting to $\IP^1\times\IP^2$ and intersection with the class $nh_1+h_2$ of $S$, or more simply by noting that $f$ is a fiber and $h$ is a section.

For $K_S$ we write $K_S=\alpha h+\beta f$ and solve for $\alpha$ and $\beta$ using the adjunction formula
\begin{equation}
f(f+K_S)=-2,\qquad h(h+K_S)=-2.
\end{equation}
We get
\begin{equation}
K_S = -2h+\left(2n-2\right)f
\end{equation}

In preparation for blowing up $S$, we need to compute its Segre class.  We have 
\begin{equation}
s(S,\widetilde{\IP^4})=c(N_{S/\widetilde{\IP^4}})^{-1}=c(T_S)c(T_{\widetilde{\IP^4}}|_S)^{-1}.
\end{equation}
We have
\begin{equation}
c(T_S)=[S]+\left(2h-(2n-2)f\right)+4p.
\end{equation}
To compute $c(T_{\widetilde{\IP^4}})$, it is most convenient to use that
$\widetilde{\IP^4}$ is a toric variety, whose six torus-invariant divisors have classes $\mathbf{H},\mathbf{H},\mathbf{H}-\mathbf{E},\mathbf{H}-\mathbf{E},\mathbf{H}-\mathbf{E},\mathbf{E}$.  We get
\begin{equation}
c(T_{\widetilde{\IP^4}})=\left(1+\mathbf{H}\right)^2\left(1+\mathbf{H}-\mathbf{E}\right)\left(1+\mathbf{E}\right).
\end{equation}

We now blow up $S$ and let $\mathbf{E}_1$ denote the exceptional divisor of this second blowup.  We then identify the class of the proper transform of $\mathbf{E}$ and denote it as
\begin{equation}
\mathbf{E}_2=\mathbf{E}-\mathbf{E}_1.
\end{equation}
Using invariance of the Segre class as we did in the $SU(2)$ case, we can compute all of the 4-fold intersections involving $\mathbf{H},\mathbf{E}_1,\mathbf{E}_2$.

We now choose our singular quintic threefold $Y$ to have multiplicity~2 along $L$, and furthermore, after blowing up as we did in the $SU(2)$ case, contains $S$.  Our Calabi-Yau $X$ is the proper transform of $Y$ after our two blowups.  We see that $X$ has class

\begin{equation}
[X]=5\mathbf{H}-2\mathbf{E}-\mathbf{E}_1.
\end{equation}
We then put
\begin{equation}
H=\mathbf{H}|_X,\ E_1=\mathbf{E}_1|_X,\  E_2=\mathbf{E}_2|_X
\end{equation}
By construction, $X$ has a resolved $A_2$ configuration over $L$.  The surface $E_1$ is a Hirzebruch surface and $E_2$ is a blown up Hirzebruch surface.  
We can then compute all of the triple products of $H,E_1,E_2$ by lifting to the blowup and multiplying by $5\mathbf{H}-2\mathbf{E}-\mathbf{E}_1$.
In particular, $N_f$ is deduced from $E_2^3$.  We omit the calculations and state the results.

For all values of $0\le i\le 3$, we get
\begin{equation}\begin{split}\label{mostint}
H^3=5,\ H^2E_1=H^2E_2=0,\ HE_1^2=HE_2^2=-2,\ HE_1E_2=1,\\ 
E_1^3=-3,\ E_2^3=8,\end{split}
\end{equation}
Since $E_1$ is a blown-up Hirzebruch surface by our general discussion and $K^2=8-N_f$ for a Hirzebruch surface blown up $N_f$ times, we conclude that $N_f=11$.  Note that $E_2$ is a Hirzebruch surface which has not been blown up, again consistent with our general discussion.

The other intersection numbers depend on $n$:
\begin{equation}\label{restint}
E_1^2 E_2 = n+3\ E_1E_2^2=-n-5.
\end{equation}

From the description of $E_1$ as a Hirzebruch surface and of $E_2$ as a blown-up Hirzebruch surface, we get $c_2(E_1)=4$ and $c_2(E_2)=15$.  From $c_2(E_i)=E_i^3+c_2\cdot E_i$ and (\ref{mostint}) we get
\begin{equation}
c_2\cdot E_1 = 18,\ c_2\cdot E_2=-4.
\end{equation}

It remains only to describe the Mori cone and K\"ahler cone.  As in the $SU(2)$ case, we choose a two-plane $Q\subset \IP^4$ containing $L$ and we again find a degree 3 curve $D\subset P\cap Y$  by (\ref{residual}).  This curve again meets $L$ in 3 points.  The factorization in (\ref{eq:su3line}) tells us that after blowing up, $n$ of these points meet $E_1$ and $3-n$ meet $E_2$.

The curves $D$ and the fiber $r_1$ of $E_1$ are again in the Mori cone.  However, unlike the $SU(2)$ case, the two components of the reducible fibers of $E_1$ are asymmetric: since a fiber $r_1$ satisfies $r_1\cdot E_2=1$, one of these two components must meet $E_2$ and the other one does not.  We let $r_1'$ be the component which intersects $E_2$.  The other component is then $r_1-r_1'$ and is disjoint from $E_2$.
In principle, we might need both $r_1'$ and $r_1-r_1'$ to generate the Mori cone. 

The intersection numbers of each of $D'\in\{D,r_1,r_2',r_2-r_2\}$ are listed as an ordered triple $[D']=(D'\cdot H,D'\cdot E_1, D'\cdot E_2)$.

\begin{equation}\label{Moricoords}
[D]= (3,n,3-n),\  [r_2]= (0,1,-2),\ [r_1']=(0,-1,1),\ [r_1-r_1']=(0,-1,0).
\end{equation}
The coordinates of $D$ can be found by an explicit geometric computation using the equation (\ref{eq:su3line}) of the singular quintic $Y$.  Alternatively, since $D$ is the complete intersection of the proper transform of two hyperplanes containing $L$, we get $D=(H-E_1-E_2)\cdot (H-E_1-E_2)$, and then the intersection numbers of $D$ with $H$, $E_1$, and $E_2$ follow readily from (\ref{mostint}) and (\ref{restint}).  

The coordinates (\ref{Moricoords}) make plain the relation $[r_1-r_1']=[r_2]+2[r_1']$.  So $[r_1-r_1']$ is not needed to span the Mori cone.

We need another curve to generate the Mori cone.  But it will be instructive to explain how we can find new curves iterative.  We content ourselves with working out the case $n=0$.

So let's assume that the Mori cone is actually spanned by $D, r_2,r_1'$.  Then dually, the K\"ahler cone would be generated by
\begin{equation}
H,\ H-E_1-E_2,\ H-2E_1-E_2.
\end{equation}
In particular $H-2E_1-E_2$.  But we compute $(H-2E_1-E_2)^3=-3<0$, contradicting Theorem~\ref{Nef}.  Thus $H-2E_1-E_2$ is not nef.

But this class does have an effective representative.  Interpreting (\ref{eq:factor}) as an equation inside the exceptional divisor $\IP^1\times\IP^2\subset\tilde{\IP^4)}$ as before, without loss of generality we can change coordinates in $\IP^2$ so that the first factor is just $x_0=0$.  Then the proper transform $H-E=H-E_1-E_2$ of $x_0=0$ contains the first Hirzebruch surface $E_1$.  So when we blow up a second time, we have to subtract $E_1$ again.  So the class of the proper transform $S$ of $x_0=0$ after both blowups is $H-2E_1-E_2$, hence that class is effective.  

Since $S$ is not nef and $S$ is a surface, a curve $C$ satisfying $S\cdot C<0$ is necessarily contained in $S$.  This gives us a strategy for finding missed curves: look for curves in $S$.  Since $|H-E_1-E_2|$ is base point free\footnote{This linear system corresponds to hyperplanes in $Y$ containing $L$.  That linear system has $L$ as a base locus, but it is immediately checked that the base locus is removed by the blowups.}, we know that we can find a representative where the intersection $S\cdot (H-E_1-E_2)$ is an effective curve $K\subset S\subset X$, hence is in the Mori cone.  Using $[S]=H-2E_1-E_2$, (\ref{mostint}) and (\ref{restint}), we get for the coordinates of $K$

\begin{equation}
[K]=(2,2,0).
\end{equation}
Comparing with (\ref{Moricoords}) and recalling that $n=0$, we see that $(1/3)[D]=(1/2)[K]+[r_1']$.  The Mori cone is generated by $K$, $r_2$, and $r_1'$.

Dually, the K\"ahler cone is generated by
\begin{equation}
H,\ H-E_1-E_2,\ 2H-2E_1-E_2.
\end{equation}
As a check, we compute $(2H-2E_1-E_2^3)=14>0$.    We have also checked  in Section~\ref{sec:u1g} that this example satisfies the physical requirements of supergravity strings.

\bibliographystyle{JHEP}

\bibliography{5dStrings}

\end{document}